\documentclass[universe,article,accept,moreauthors,pdftex]{Definitions/mdpi} 

\usepackage{amsmath,amssymb}
\usepackage{epsfig}

\newcommand{\cd}{\! \cdot \!}

\newcommand{\ba}{\begin{eqnarray}}
\newcommand{\ea}{\end{eqnarray}}

\newcommand{\be}{\begin{equation}}
\newcommand{\ee}{\end{equation}}
\newcommand{\pa}{\partial}

\firstpage{1} 
\pubvolume{7}
\issuenum{3}
\articlenumber{59}
\pubyear{2021}
\copyrightyear{2021}
\externaleditor{Academic Editor: {Nicolas Chamel}} 
\datereceived{21 January 2021} 
\dateaccepted{27 February 2021 } 
\datepublished{5 March 2021} 
\hreflink{https://doi.org/10.3390/ \linebreak universe7030059} 

\pdfoutput=1

\Title{Transport Properties of Superfluid Phonons in Neutron Stars}
\TitleCitation{Transport Properties of Superfluid Phonons in Neutron Stars}

\Author{{Cristina Manuel} 
 $^{1,2}$\orcidA{} and Laura Tolos $^{1,2,3,4,}$*\orcidB{}}

\AuthorNames{Cristina Manuel and Laura Tolos}
\AuthorCitation{Manuel, C.; Tolos, L.}
\address{%
$^{1}$ \quad Institute of Space Sciences (ICE, CSIC), Campus UAB, Carrer de Can Magrans, 08193 Barcelona, Spain; {cmanuel@ice.csic.es}\\ 
$^{2}$ \quad Institut d'Estudis Espacials de Catalunya (IEEC), 08034 Barcelona, Spain \\
$^{3}$ \quad Frankfurt Institute for Advanced Studies, Ruth-Moufang-Str. 1, 60438 Frankfurt am Main, Germany \\
$^{4}$ \quad Faculty of Science and Technology, University of Stavanger, 4036 Stavanger, Norway}

\corres{Correspondence: tolos@ice.csic.es}

\abstract{ We review the effective field theory associated with the superfluid phonons that we use for the study of transport properties in the core of superfluid neutrons stars in their low temperature regime. We then discuss
 the shear and bulk viscosities together with the thermal conductivity coming from the collisions of superfluid phonons in neutron stars. With regard to shear, bulk, and thermal transport coefficients, the phonon collisional processes are obtained in terms of the equation of state and the superfluid gap. We compare the shear coefficient due to the interaction among superfluid phonons with other dominant processes in neutron stars, such as electron collisions. We also analyze the possible consequences for the r-mode instability in neutron stars. As for the bulk viscosities, we determine that phonon collisions contribute decisively to the bulk viscosities inside neutron stars. For the thermal conductivity resulting from phonon collisions, we find that it is temperature independent well below the transition temperature. We also obtain that the thermal conductivity due to superfluid phonons dominates over the one resulting from electron-muon interactions once phonons are in the hydrodynamic regime. As the phonons couple to the $Z$ electroweak gauge boson, we estimate the associated neutrino emissivity. We also briefly comment on how the superfluid phonon interactions are modified in the presence of a gravitational field or in a moving background.
}


\begin{document}

\section{Introduction}

Superfluidity is a property of quantum liquids related to the existence of low energy excitations
that satisfy the so-called Landau criterion~\cite{Landau:1941vsj}
\be
{\rm Min} \frac{\epsilon(p)}{p} \neq 0 \ ,
\ee
where $\epsilon(p)$ is the dispersion law of the excitation. The lowest energy modes are usually called superfluid phonons, and they dominate
the low temperature properties of the superfluid.
 While superfluidity was first discovered in a bosonic system,
it was soon realized that it could occur in a fermionic system as well, as a consequence 
of Cooper's theorem.

Migdal's observation~\cite{Migdal} that the superfluidity of neutron matter may occur in the core of compact stars has captured much interest over the years.
The presence of superfluidity inside neutron stars would affect different neutron star phenomena, such as its cooling, rotational properties, pulsar glitches, or the hydrodynamic and oscillation modes, and thus have several observational consequences. For recent reviews on superfluidity in neutron stars, we refer the reader to~\cite{Sedrakian:2018ydt,Page:2013hxa}.

At low temperatures, neutron matter superfluidity originates from the presence of a quantum condensate, due to neutron pairing. This condensate breaks the global 
$U(1)$ symmetry associated with baryon number conservation and gives rise to the existence of the low energy modes, the superfluid phonons. At very low temperatures, they dominate the thermal corrections to the thermodynamical and hydrodynamic properties of the superfluid. More precisely, the superfluid phonon contribution could be relevant for the determination of the transport properties of neutron stars, that is the shear and bulk viscosities, as well as the thermal conductivity.

In this manuscript, we review the progress on the transport properties of superfluid phonons in the core of the star. In the crust of the star, neutrons also pair, but the low
energy effective field theory includes other low energy modes, which are absent in the core~\cite{Page:2012zt}.~We give an overview of the superfluid phonon contribution to the shear and bulk viscosities together with the thermal conductivity in the core of neutron stars, as discussed in~\cite{Manuel:2011ed,Manuel:2012rd,Manuel:2013bwa,Manuel:2014kqa}. So as to calculate the phonon contribution, it is important to determine the relevant collisions involving phonons that are responsible for the transport phenomena. By using the universal character of the effective field theory (EFT) at leading order, we can obtain a very general formulation so as to determine the leading phonon interactions, which can be fixed by the equation of state (EoS) and the superfluid gap of neutron matter~\cite{Son:2002zn,Son:2005rv}. We also discuss neutrino emission by superfluid phonons. Even if these couple to the $Z$ electroweak gauge boson, this possible channel of cooling of the star is very much suppressed, as we will show.

The paper is organized as follows. In Section~\ref{sec:eft}, we review the EFT that describes the phonon self-interactions in a superfluid, whereas in Section~\ref{sec:eos-gap}, we 
present the specific EoS for $\beta$-stable nuclear matter, as well as the superfluid gap that will be used for the computation of the transport coefficients. In Section~\ref{sec:shear}, 
 we study the shear viscosity in the hydrodynamic (Section~\ref{sub-hydroshear}) and ballistic and transient regimes (Section~\ref{ball-sec}). In \mbox{Section~\ref{sec:bulk}}, we show our results for {the} bulk viscosities, and in \mbox{Section~\ref{sec:r-modes}}, we analyze the r-mode instability window using the shear damping mechanism. Then, in Section~\ref{sec:thermal}, we discuss the thermal conductivity, and in Section~\ref{sec:neutrino}, we comment on neutrino emission due to superfluid phonons. We finalize our paper with Section~\ref{sec:gravity}, where we present a discussion on how the superfluid phonon EFT changes in the presence of a gravitational field or by considering that the superfluid medium is not at rest. A summary is given in 
 Section~\ref{sec:summary}. We use natural units $\hbar = c = k_B=1$ in this work.

\section{Effective Field Theory and the Superfluid Phonon}
\label{sec:eft}

EFT techniques allow writing down the effective Lagrangian for superfluid phonons. A superfluid phonon is the Goldstone mode related to the spontaneous symmetry breaking of the particle number conservation. The effective Lagrangian is constructed as an expansion in the derivatives of the superfluid phonon field, each of the terms restricted by the allowed symmetry. The coefficients of each term in the expansion can be determined from the microscopic theory using a matching procedure.
An important observation was made in~\cite{Son:2002zn,Son:2005rv}, as it was pointed out that the EoS determines completely the leading-order (LO) effective Lagrangian for phonons. In particular, the LO Lagrangian for a non-relativistic system reads: 
\begin{eqnarray}
\label{LO-Lagran}
\mathcal{L}_{\rm LO} &=&P (X) \ , \nonumber \\
 X &=& \mu-\partial_t\varphi-\frac{({\bf \nabla}\varphi)^2}{2m} \ ,
\end{eqnarray}
where $P(\mu)$ and $\mu$ are the pressure and chemical potential, respectively, of the system at $T=0$, whereas
$\varphi$ is the phonon field (the phase of the fermionic condensate) and $m$ is the mass of the
particles that form a condensate. Note that, after a Legendre
transformation, the associated Hamiltonian can be obtained, having the same form as the one used by Landau for
the phonon self-interactions of $^4$He~\cite{Son:2005rv,Khalatnikov:106134}.

The Lagrangian for the phonon field is determined after performing a Taylor expansion of the pressure and rescaling the phonon field: 
\be
\label{normalization}
\varphi = \frac{\phi}{\sqrt{ \frac{\pa^2 P}{\pa \mu^2 }}} ,
\ee
 in order to canonically normalize the kinetic term as:
\begin{eqnarray}
\label{comlag}
\mathcal{L}_{\rm LO}&&=\frac{1}{2}\left((\partial_t\phi)^2-v^2_{\rm ph}({\bf \nabla}\phi)^2\right) \nonumber \\
&&-g\left((\partial_t \phi)^3-3\eta_g \,\partial_t \phi({\bf \nabla}\phi)^2 \right) \nonumber \\
&&+\lambda\left((\partial_t\phi)^4-\eta_{\lambda,1} (\partial_t\phi)^2({\bf \nabla}\phi)^2+\eta_{\lambda, 2}({\bf \nabla}\phi)^4\right)
+ \cdots
\end{eqnarray}

In~\cite{Escobedo:2010uv} the phonon self-couplings of Equation~(\ref{comlag}) were obtained 
as different ratios of the derivatives of the pressure with respect to the chemical potential. Similarly, we can express them in terms of the speed of sound at $T=0$ and derivatives of the speed of sound with respect to the mass density. 

In particular, the speed of sound at $T=0$ is given by: 
\begin{equation}
\label{phspeed}
v_{\rm ph}= \sqrt{\frac{\partial P}{\partial {\rho}} } \equiv c_s \ ,
\end{equation}
\textls[-25]{where ${\rho}$ is the mass density. The three and four phonon self-coupling constants are obtained as:}
\be
\begin{split}
&g=\frac{1-2 u}{6 c_s \sqrt{\rho}}\,,\qquad \eta_g=\frac{c^2_s}{1-2 u}\,,\qquad \lambda=\frac{1-2 u(4-5u)-2 w \rho}{24c^2_s\rho}\,,\\
&\eta_{\lambda\,,1}=\frac{6c^2_s(1-2 u)}{1-2u(4-5 u)-2w\rho}\,,\qquad \eta_{\lambda\,,2}=\frac{3c^4_s}{1-2u(4-5 u)-2w\rho} \ ,
\label{eq:relations}
\end{split}
\ee
with:
\be
u=\frac{\rho}{c_s}\frac{\partial c_s}{\partial\rho}\,, \quad w=\frac{\rho}{c_s}\frac{\partial^2 c_s}{\partial\rho^2}\,.
\label{precoup}
\ee
Note that the dispersion law coming from this Lagrangian at the tree level is exactly $E_p = c_s p $, so that phonons move at the speed of sound.

The formulation of the superfluid phonon EFT just presented is universal and valid for different superfluid systems, either bosonic or fermionic, and no matter
if those systems are weakly coupled or not. Thus, with the same formulation, we can describe the superfluid phonons of either superfluid $^4$He, cold Fermi gases at unitarity, or the superfluid neutron matter inside neutron stars. With very minor modifications, relativistic superfluids, such as the color-flavor locked quark matter phase, can be also described within the same methods~\cite{Son:2002zn}. 
With the knowledge of the EoS, many of the low temperature properties of the superfluid, mainly those associated with transport, can be immediately deduced. For astrophysical applications,
our treatment is very convenient, as we could provide the values associated with the transport phenomena involving superfluid phonons in terms of 
different EoSs.
 We ignore here the impact on the superfluid phonon EFT of the entrainment effects that couple the neutron and proton fractions in the star. At most, these could slightly modify the explicit values of the coupling constants displayed in
Equation~(\ref{eq:relations}). Note that Equation~(\ref{comlag}) is the most general Lagrangian with respect to the symmetries of the system.

It also is possible to construct the next-to-leading order (NLO) Lagrangian in a derivative expansion. As seen in~\cite{Son:2005rv},
this reads:
\be
\mathcal{L}_{\rm NLO} = \pa_i X \pa_i X f_1(X) + (\Delta^2 \theta)^2 f_2 (x) \ ,
\ee
where $\theta = \mu t - \varphi$ and $f_1$ and $f_2$ are arbitrary functions. Unfortunately, there is a not a simple way of finding the values of
these functions. 
For the cold Fermi gas at unitarity, the NLO Lagrangian can be determined up to two arbitrary constants, just demanding invariance under scale transformations. However, this is not a symmetry in superfluid neutron matter, for example.

The NLO Lagrangian is relevant to study different corrections to the different scattering rates among superfluid phonons, which will be typically
minor corrections. However, it is important for determining the corrections to the phonon dispersion law. The NLO phonon dispersion law is given by:
\be
\label{NLOdisp-law}
E_P = c_s p ( 1 + \gamma p^2) \ , 
\ee
with:
\be
\gamma = - \frac{1}{ \frac{\pa^2 P}{\pa \mu^2}} \left( f_1(\mu) + \frac{f_2(\mu)}{c_s^2} \right) \ .
\ee
The sign of $\gamma$ determines whether the decay of one phonon into two (or more) phonons is kinematically allowed.
More particularly, with the LO Lagrangian, energy and momentum conservation allows for the decay of one phonon into two, if the three 
 involved phonons are collinear. If we take into account the NLO corrections, energy and momentum conservation impose that the small angle
 $\delta \theta$ among the
two resulting phonons, with momenta $p_b$ and $p_c$, respectively, is:
\be
\delta \theta = \sqrt{ 6 \gamma} (p_b+p_c) \ .
\ee
Thus, the process is only kinematically allowed if $\gamma > 0$.

In the computation of the transport properties of a superfluid associated with the phonons, it becomes thus essential to determine the value of
$\gamma$, so as to know the possible scattering processes that contribute to a particular transport coefficient. 
For the astrophysical applications, we should have in mind that the value of $\gamma$ was computed assuming that neutron pairing is in a $^1S_0$ channel, treating the neutrons as a weakly coupled system, resulting in~\cite{Manuel:2014kqa}:
 \be
\gamma=-\ \frac{v_F^2}{45 \Delta^2} , 
\ee
with $v_F$ being the Fermi velocity and $\Delta$ the gap value in the $^1S_0$ phase. In the following, we assume that $\gamma$ takes this same value in the $^3P_2$ phase, where $\Delta$ is the angular averaged gap in that phase. Therefore, taking into account that $\gamma <0$, the first allowed phonon scattering processes are binary collisions. It might be interesting to compute the value of $\gamma$ for more realistic neutron-neutron interactions. Furthermore, we note that the value of the gap is temperature dependent, vanishing close to critical temperature $T_c$. However, we assumed a temperature independent value for the determination of $\gamma$.

We ignore in this manuscript the interaction of the superfluid phonons with the electrons in the core of the star, as we estimated that this interaction cannot
give the leading effect in the transport coefficients we consider in this manuscript; see~\cite{Manuel:2012rd}.

 Let us point out that when triplet pairing occurs, apart from the superfluid phonons, the angulons, which are the Goldstone modes associated with the spontaneous symmetry
breaking of the rotational symmetry, should be considered as well, and in principle, they should give a contribution to the transport coefficients.
An effective field theory for these modes was developed in~\cite{Bedaque:2003wj}, where essentially only the interactions of the angulons with the electroweak gauge boson were considered. It would be interesting to consider all the possible interactions of these modes with other low energy modes as well to consider their impact on the hydrodynamics, which should then reflect the breaking of the rotational symmetry. This is an interesting point that deserves consideration, as to date, the hydrodynamic equations used to describe superfluid neutron stars do not consider such an~effect.


\section{Equation of State and the Gap of Neutron Matter in Superfluid Neutron Stars}
\label{sec:eos-gap}

\begin{figure}[H]
\includegraphics[width=0.6\textwidth]{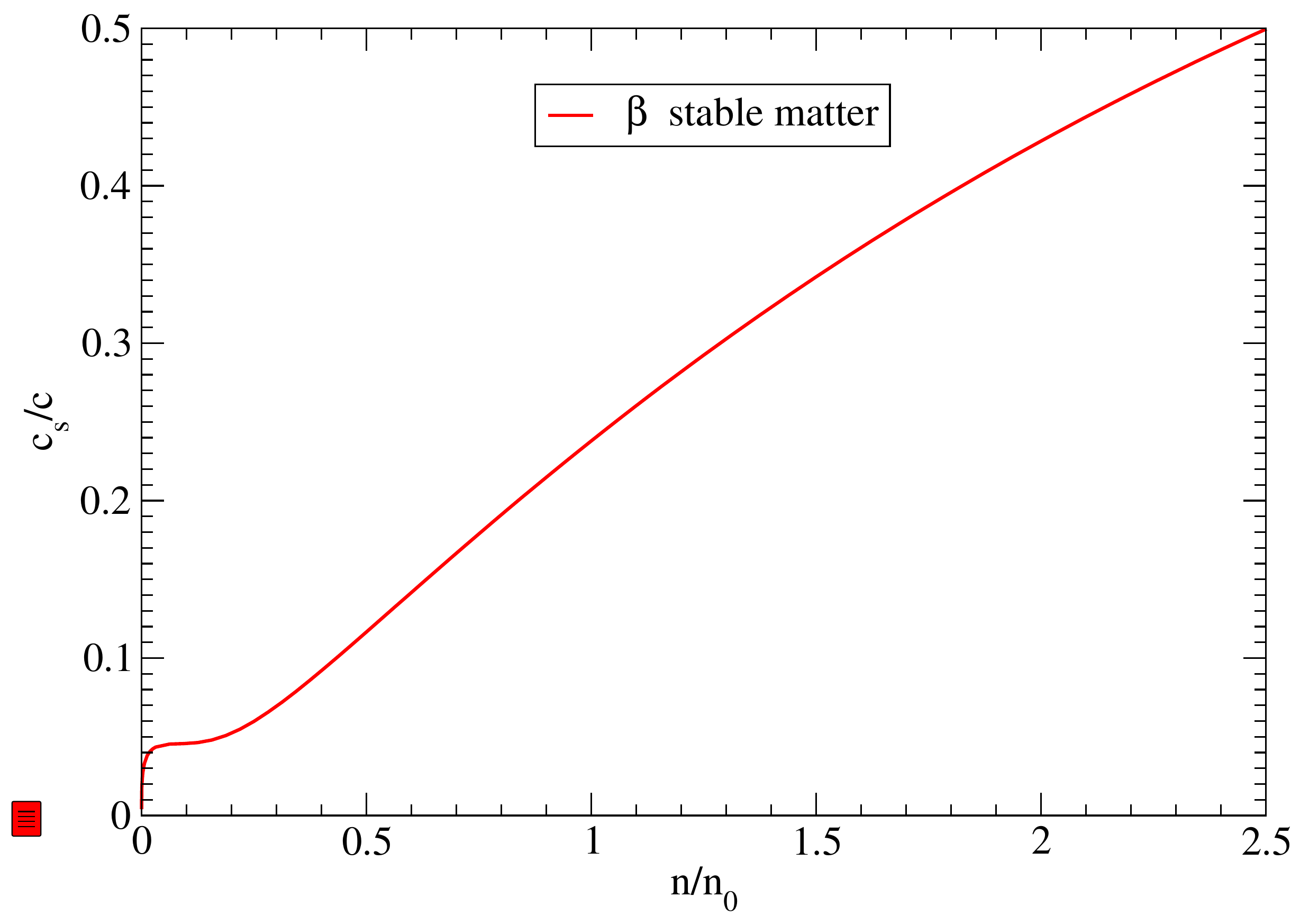}
\includegraphics[width=0.6\textwidth]{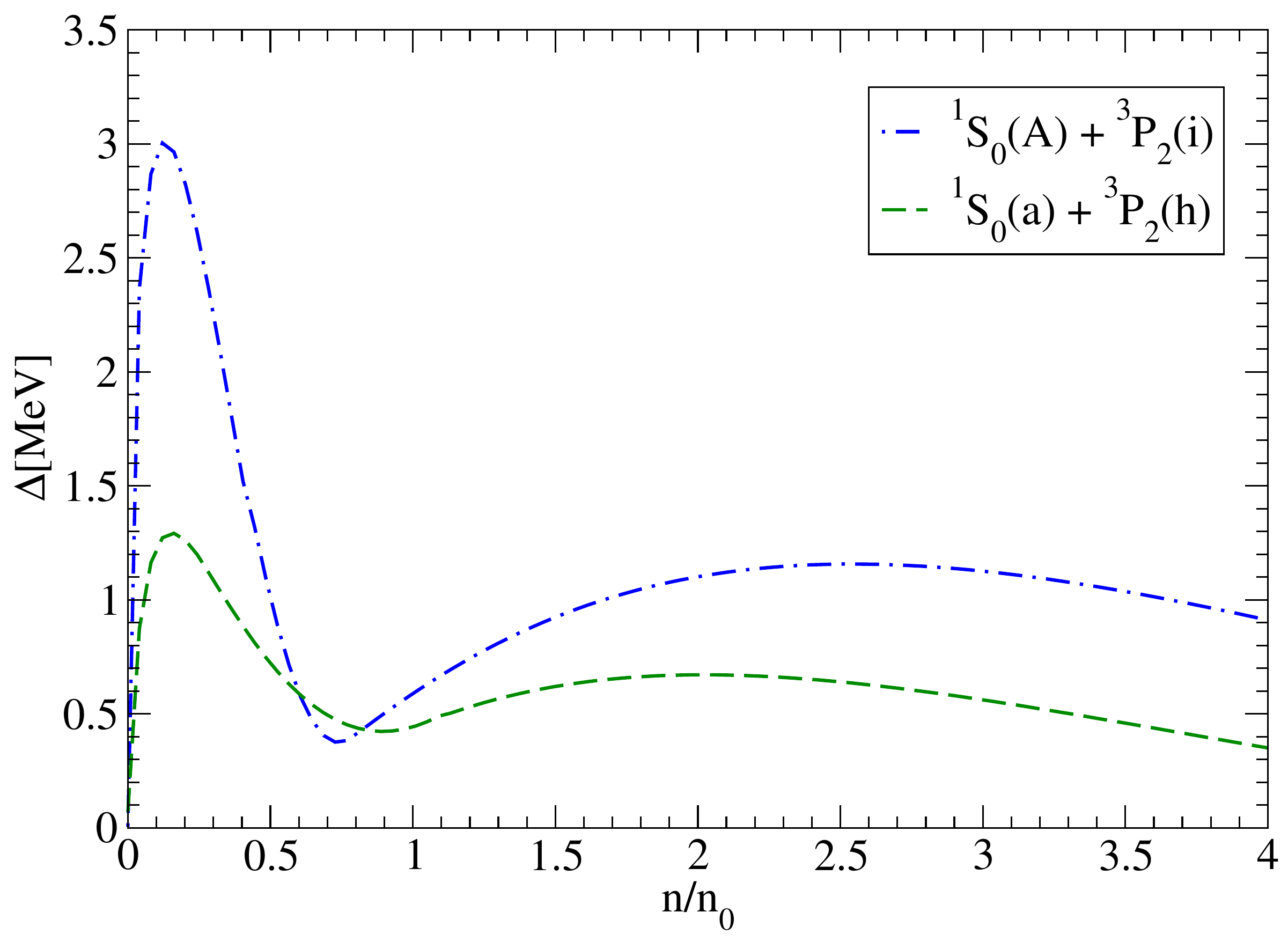}
\caption{{Upper panel}: $c_s/c$ for $\beta$-stable nuclear matter as a function of particle density. Lower panel: The combination of $^1S_0$ and angle-averaged $^3P_2$ neutron gaps as a function of density for two different gap cases. {Figures} from~\cite{Manuel:2011ed,Manuel:2013bwa,Tolos:2014wla}. Adapted from \cite{Tolos:2014wla} with the {permission of AIP Publishing}.}
\label{speeds-gap}
\end{figure}

The EoS in neutron stars is needed so as to determine not only the speed of sound at $T=0$, but also the phonon self-couplings. 
The theoretical description of matter in neutron stars is a complicated task given that the EoS spans over a wide range of densities, temperatures, and isospin asymmetries (see, for example, the recent review~\cite{Tolos:2020aln}). More precisely, inside the core of neutron stars, nuclear matter can be described using different theoretical many-body schemes. Those are usually divided between microscopic ab-initio approaches and phenomenological models. Whereas microscopic ab-initio approaches obtain the EoS solving the many-body problem from two-body and three-body interactions, which are fitted to experimental data on scattering and finite nuclei, phenomenological models rely on density-dependent interactions adjusted to nuclear observables and neutron star predictions. 

Among the first ones, one finds the well-known variational method of Akmal, Pandharipande, and Ravenhall (APR)~\cite{Akmal:1998cf}, which is commonly used as a benchmark for other EoSs. Heiselberg and Hjorth-Jensen parametrized the APR EoS in a simple and handleable form in~\cite{Heiselberg:1999mq}. In this paper, we make use of this EoS. Note that the effect of neutron pairing in the EoS is not taken into account as $\Delta/\mu << 1$. 

In the upper panel of Figure~\ref{speeds-gap}, we show the speed of sound versus the speed of light $c_s/c$ in $\beta$-stable nuclear matter as a function of the density. We observe 
that relativistic effects become important for densities of the order of 1.5--2 $n_0$, as the speed of sound increases monotonically with density. 





With regard to the value of the gap of superfluid matter, in the lower panel of \mbox{Figure~\ref{speeds-gap}}, we show the two gap models we will use in this work as a function of the density. These extreme cases have been chosen so as to account for the model dependence of our results. These models are named $^1S_0(A)$$+$$^3P_2(i)$ and $^1S_0(a)$$+$$^3P_2(h)$, following the notation of~\cite{Andersson:2004aa}, and take into account a wide range of gap values inside the core of neutron stars. 

The $^1S_0(A)$$+$$^3P_2(i)$ scheme results from the combination, on the one hand, of the $^1S_0$ neutron gap coming from the parametrization $A$ of Table I in~\cite{Andersson:2004aa}, which takes into account the {BCS  approach} of different nuclear interactions with a maximum gap of approximately 3 MeV at $p_F \approx 0.85~{\rm fm}^{-1}$, and, on the other hand, the anisotropic $^3P_2$ neutron gap from the parameterization $i$ (strong neutron superfluidity in the core) of Table I in~\cite{Andersson:2004aa}, a model-dependent result that goes beyond BCS theory. 
 Note, however, that those maximum values for the s-wave and p-wave gaps are not too realistic as they are obtained from nucleon-nucleon bare interactions. The inclusion of many-body corrections will reduce those values, as shown in~\cite{Sedrakian:2018ydt}.
{The $^1S_0(a)$$+$$^3P_2(h)$ model} goes beyond BCS for the $^1S_0$ neutron gap as it incorporates medium polarization effects (parametrization $a$), reducing the maximum value to 1 MeV, whereas the $^3P_2$ neutron gap comes from the parametrization $h$ (strong neutron superfluidity) with a maximum value of about 0.5 MeV. For both models, the transition temperatures from the superfluid to the normal phase are $T_c \sim 1/2 \Delta \gtrsim 0.25 \times 10^8$ K. Note that for the values of the gap of 0.5 MeV, the transition temperature from the superfluid to the normal phase is $T_c \sim (1/2) \Delta \gtrsim 2.5 \times 10^9$ K.

\section{The Shear Viscosity of Superfluid Phonons}
\label{sec:shear}

\subsection{Hydrodynamic Regime}
\label{sub-hydroshear}

The shear viscosity $\eta$ emerges as a dissipative term in the energy-momentum tensor $T_{ij}$. If one performs small deviations from equilibrium, one finds that:
\be
\label{shear-stress}
 \delta T_{ij}=- \eta \tilde V_{ij} \equiv - \eta\left( \partial_i V_j+ \partial_j V_i -\frac 23 \delta_{ij} \nabla \cd {\bf V} \right) \ , 
 \ee
${\bf V}$ being the fluid velocity of the normal component of the system.

\begin{figure}[t]
\includegraphics[width=0.7\textwidth, height=0.2\textwidth]{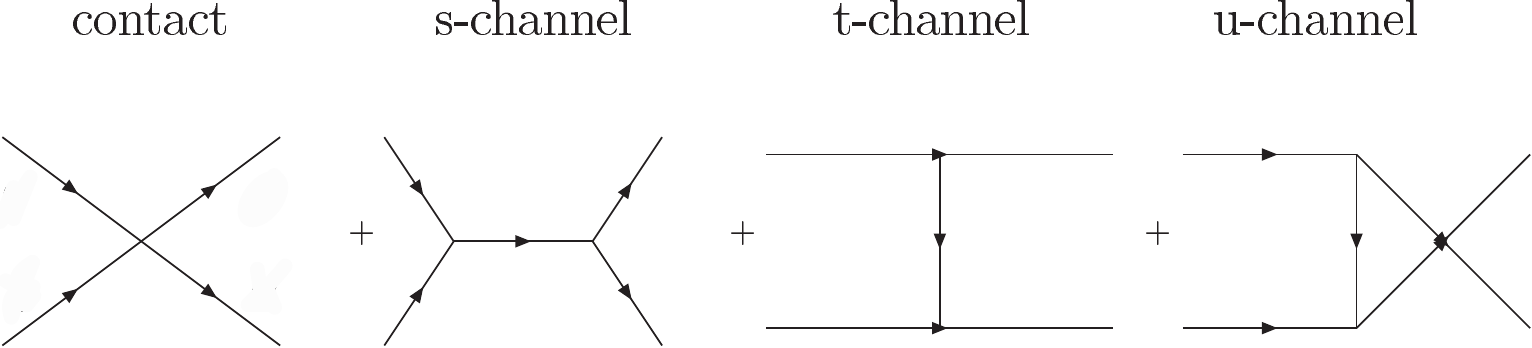}
 \caption{The $2\leftrightarrow 2$ phonon scattering processes contributing to the shear viscosity. {Figure adapted from}~\cite{Manuel:2011ed}.}
 \label{feyndiags}
\end{figure}

The superfluid phonon contribution to $T_{ij}$ is given by:
 \be
 T_{ij}= c_s^2 \int \frac{d^3 p}{(2 \pi)^3} \frac{ p_i p_j}{E_p} f(p,x) \ , 
 \ee
where $f$ is the phonon distribution function that obeys the Boltzmann equation~\cite{Khalatnikov:106134}: 
 \be
 \label{transport}
 \frac{df}{dt} = \frac{\partial
f}{\partial t}+ \frac{\partial E_p}{\partial \bf p} \cdot \nabla f= C[f] \ ,
\ee
assumed to be in the superfluid rest frame, and $C[f]$ is the collision term. As mentioned before, in order to obtain the shear viscosity due to superfluid phonons, it is enough to consider binary collisions in the collision term. Those are depicted in Figure~\ref{feyndiags}. Note that we keep the phonon dispersion law at LO for the computation of the shear viscosity.

The shear viscosity is calculated by considering small departures from equilibrium to the phonon distribution function and, afterwards, linearizing the corresponding transport equation~\cite{Manuel:2011ed}. Then, we use variational methods so as to solve the transport equation~\cite{Manuel:2004iv,Alford:2009jm,Rupak:2007vp}. The expression for the shear viscosity then reads~\cite{Manuel:2011ed}:
\be
\eta=\left( \frac{2 \pi}{15} \right)^4 \frac{T^8}{c_s^8} \frac{1}{M} \ ,
\ee 
where $M$ is a multidimensional integral that takes into account the thermally weighted scattering matrix for phonons.

\begin{figure}[H]
\includegraphics[width=0.55\textwidth]{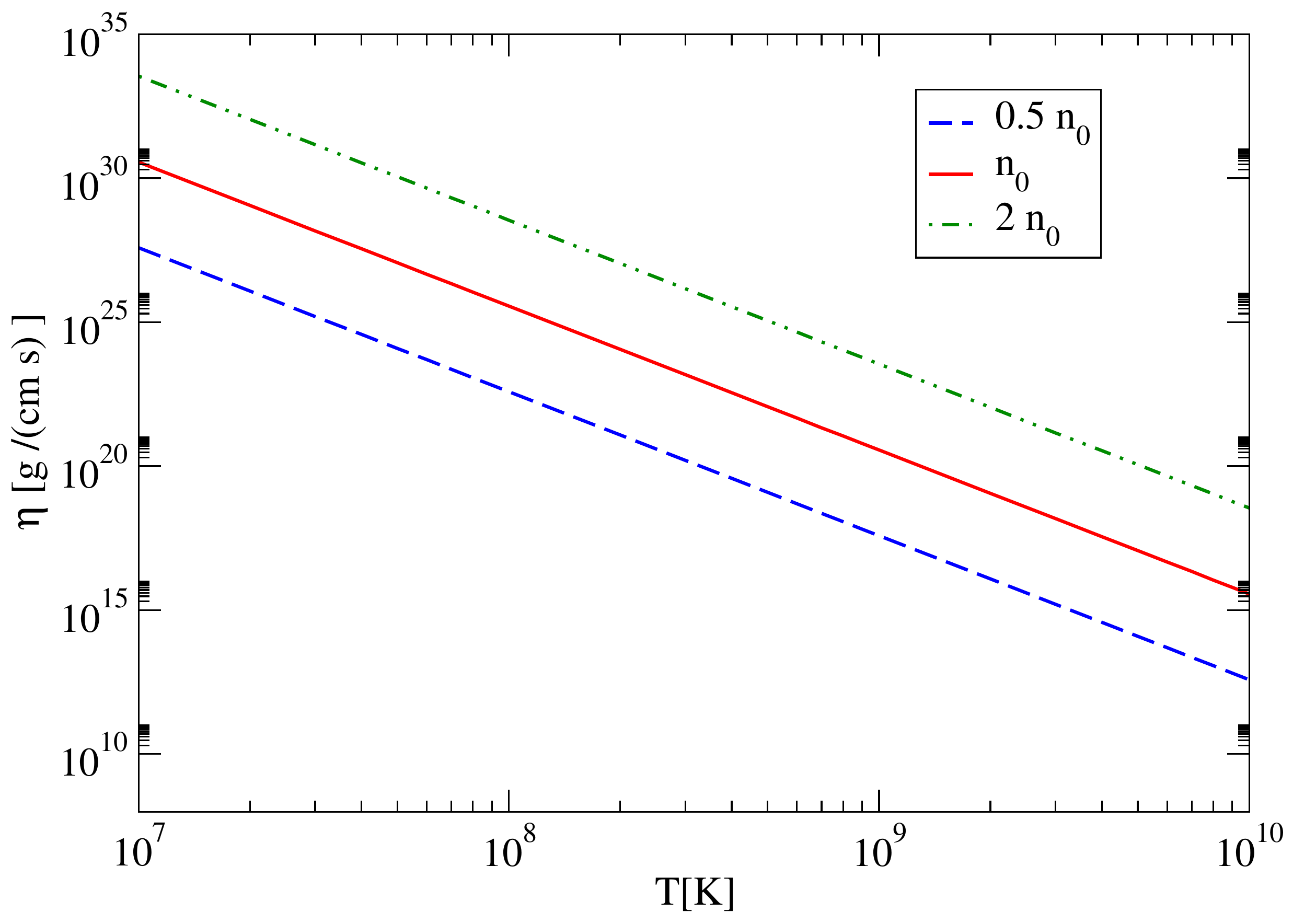}
\includegraphics[width=0.55\textwidth]{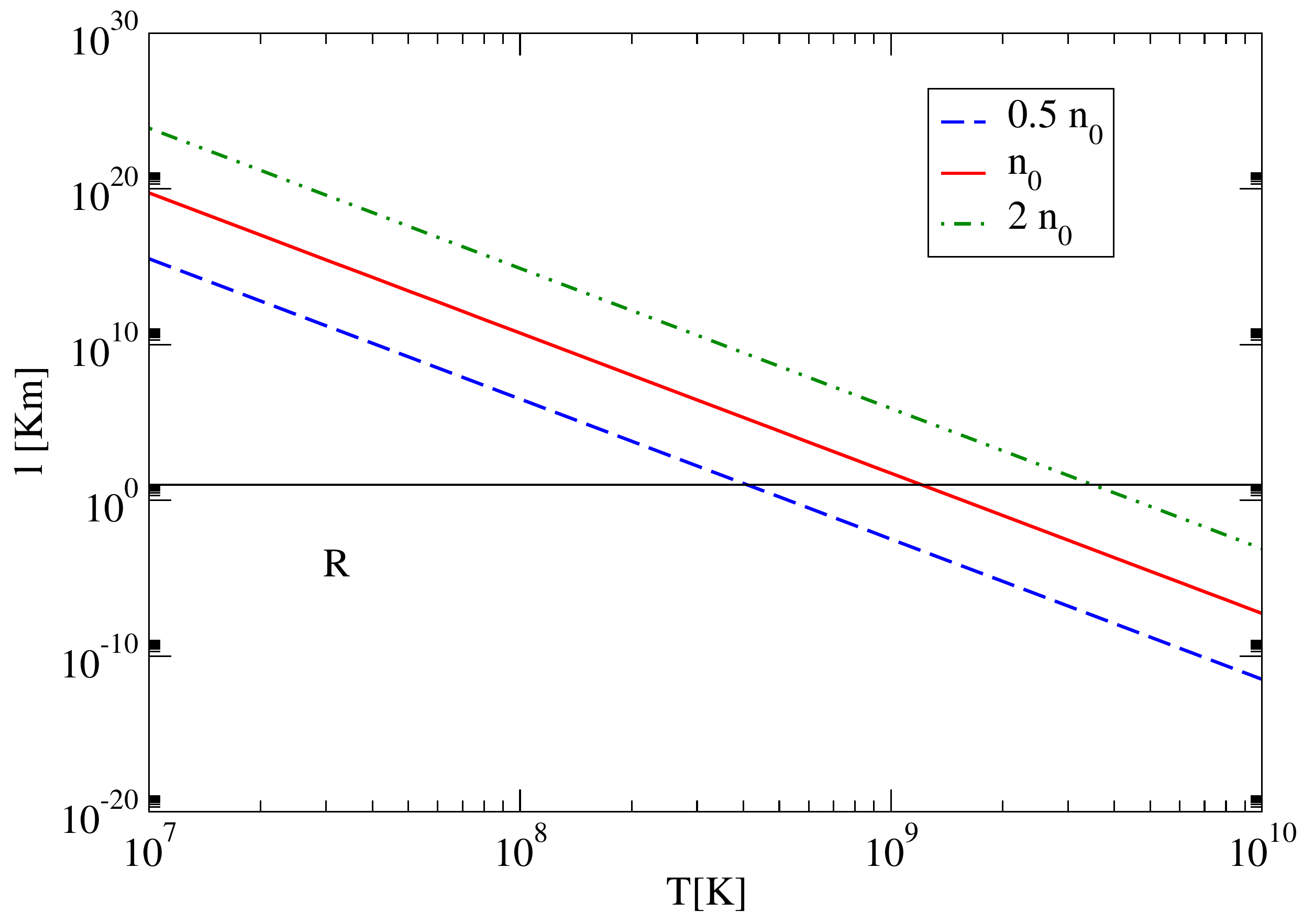}
\caption{{\textbf{Upper panel}}: Shear viscosity in $\beta$-stable neutron matter for three densities as a function of temperature. \textbf{Lower panel}: Phonon mean free path in $\beta$-stable matter as function of temperature for three densities. The horizontal line indicates the radius of the neutron star of $R = 10 \ {\rm km}$. Figures adapted from~\cite{Manuel:2011ed,Tolos:2014wla}. Adapted from \cite{Tolos:2014wla} with the permission of {AIP Publishing}.} 
\label{fig:shear} 
\end{figure}

In the upper panel of Figure~\ref{fig:shear}, we display the shear viscosity due to the binary phonon collisions as a function of the temperature for different densities. We find that $ \eta \propto 1/T^5$. In fact, this is a universal feature that occurs in other superfluid systems, such as ${\rm^4 He}$~\cite{Khalatnikov:106134} or superfluid cold atoms at unitarity~\cite{Rupak:2007vp,Mannarelli:2012eg}. The value of $\eta$ is, however, determined by the choice of the EoS.
 
The next question is at which densities and temperatures the hydrodynamic regime can be reached. The hydrodynamic regime is only achieved when the mean free path (mfp) is smaller than the typical macroscopic length of the system, in our case the radius of the star. Therefore, in the lower panel of Figure~\ref{fig:shear}, we present the mfp of phonons for different densities as a function of temperature, also showing the radius of the star, which we take as 10 km. The mfp $l$ results from the calculation of $\eta$ in~\cite{Alford:2009jm}:
\ba
l=\frac{\eta}{n <p>} ,
\label{eq1}
\ea
\ba
<p>=2.7 \frac{T}{c_s} \ , \qquad
n=\int \frac{d^3p}{(2\pi)^3} f_p=\xi(3) \frac{T^3}{\pi^2 c_s^3} .
\label{eq2}
\ea
with $<p>$ the thermal average momentum and $n$ the phonon density. 

We find that a hydrodynamic description starts being questionable for temperatures below $T \sim 10^9$ K. Note that the critical temperature for the phase transition to the normal phase is $T_c \sim 10^{10}$ K. The value for the critical temperature, as well as the values for the phonon mfp depend on the model for the EoS.

\subsection{Ballistic and Transient Regime}
\label{ball-sec}

At sufficiently low temperatures, the phonon contribution to $\eta$ diverges. Nevertheless, this result is unphysical as it does not consider finite size effects that prevent viscosity from increasing at low temperatures indefinitely. As shown in~\cite{Manuel:2011ed} and seen in Figure~\ref{fig:shear}, the phonon mean free path exceeds the size of the star at sufficiently low $T$. We can define
the viscous mean free path as:
\be
\label{bulkvis}
\eta_{\rm bulk} = \frac 15 \rho_{\rm ph} c_s l_{\rm ph} \ ,
\ee
with $\eta_{\rm bulk}$ the phonon contribution to the shear viscosity in the hydrodynamic regime, whereas
 $\rho_{\rm ph} = \frac{2 \pi^2 T^4}{45 c_s^5}$ is the phonon contribution to the mass density. In order to be in the hydrodynamic regime, the Knudsen number ($K_n$) has to be small. In our case, the Knudsen number is defined as the ratio of the mean free path versus the radius of the core of the star $R_c$, so $K_n = \frac{l_{\rm ph}}{R_c}$. As seen in Figure~\ref{fig:shear} (lower panel), the Knudsen number is $K_n \lesssim 1$ for $T \gtrsim 5 \times 10^8-10^9$ K, where the precise value of $T$ depends
on the value of the nucleon particle density. In the case of $K_n > 1$, one has to write a Boltzmann equation to describe the phonon dynamics, as a hydrodynamic description of the phonons is not possible.

For $K_n > 1$, that is when the phonon mean free path exceeds the size of the superfluid core of the star, the phonon transport properties are mainly governed by their interactions with the boundary of the superfluid region, that is the crust of the star. For simplicity, we assume that the crust is not in a superfluid phase, as done in~\cite{Bildsten:1999zn,Glampedakis:2006mn}.~If the motion of the superfluid phonons is confined to the core, then phonons reaching the crust can be absorbed or scattered back. If the phonons are diffused, they will exert a shear stress on the boundary. In this case, phonons are in a ballistic regime, where no hydrodynamic description of the phonon behavior is possible. Nevertheless, it has been experimentally observed that for $^4$He at a very low temperature, ballistic phonons can still efficiently damp the movement of immersed objects~\cite{Eselson,Niemetz2004,Zadorozhko}. This damping can still be described by a ballistic 
shear viscosity coefficient determined as:
\be\label{etaball-He4}
\eta_{\rm ball} \equiv \frac{1}{5} \rho_{\rm ph}\chi c_s d\,,
\ee
with $d$ the typical size of the object and $\chi$ measuring the probability that the phonons are diffused at the boundary. Indeed, this expression is in excellent agreement with the experimental data for large Knudsen numbers~\cite{Zadorozhko}.

In the intermediate regime of Knudsen numbers, one can effectively obtain the shear viscosity as:
\be
\eta_{\rm eff} = \left( \eta_{\rm bulk}^{-1} + \eta_{\rm ball}^{-1}\right)^{-1} \, .
\label{eff}
\ee
The justification of this formula comes from the fact that the shear viscosity is proportional to a collisional relaxation time and that the total relaxation frequency is the sum of the partial relaxation frequencies, considering different processes occurring independently with different relaxation times. This type of formula has also been used for the description of the low $T$ behavior of the shear viscosity due to phonons in a cold Fermi atomic gas in the unitarity limit~\cite{Mannarelli:2012eg}.

As discussed in~\cite{Manuel:2012rd}, we use Equation~(\ref{eff}) to consider the phonon contribution to $\eta$ in the core of the star, where $d$ is the value of the radius of the core. The radius of the core is fixed by the transition between the crust and the core and corresponds to $n \sim 0.5 \ n_0$. As for the value of $\chi$, we assume that $\chi \sim 1$. For $\eta_{\rm bulk}$, we use the values of the shear viscosity of~\cite{Manuel:2011ed} in the hydrodynamic regime and shown in Figure~\ref{fig:shear} (upper panel). We note that $\eta_{\rm eff}$ is dominated by the smaller contribution among $\eta_{\rm bulk}$ and $\eta_{\rm ball}$, so it describes properly the different hydrodynamic, transition, and ballistic regimes.

\begin{figure}[H]
\includegraphics[width=0.63\textwidth]{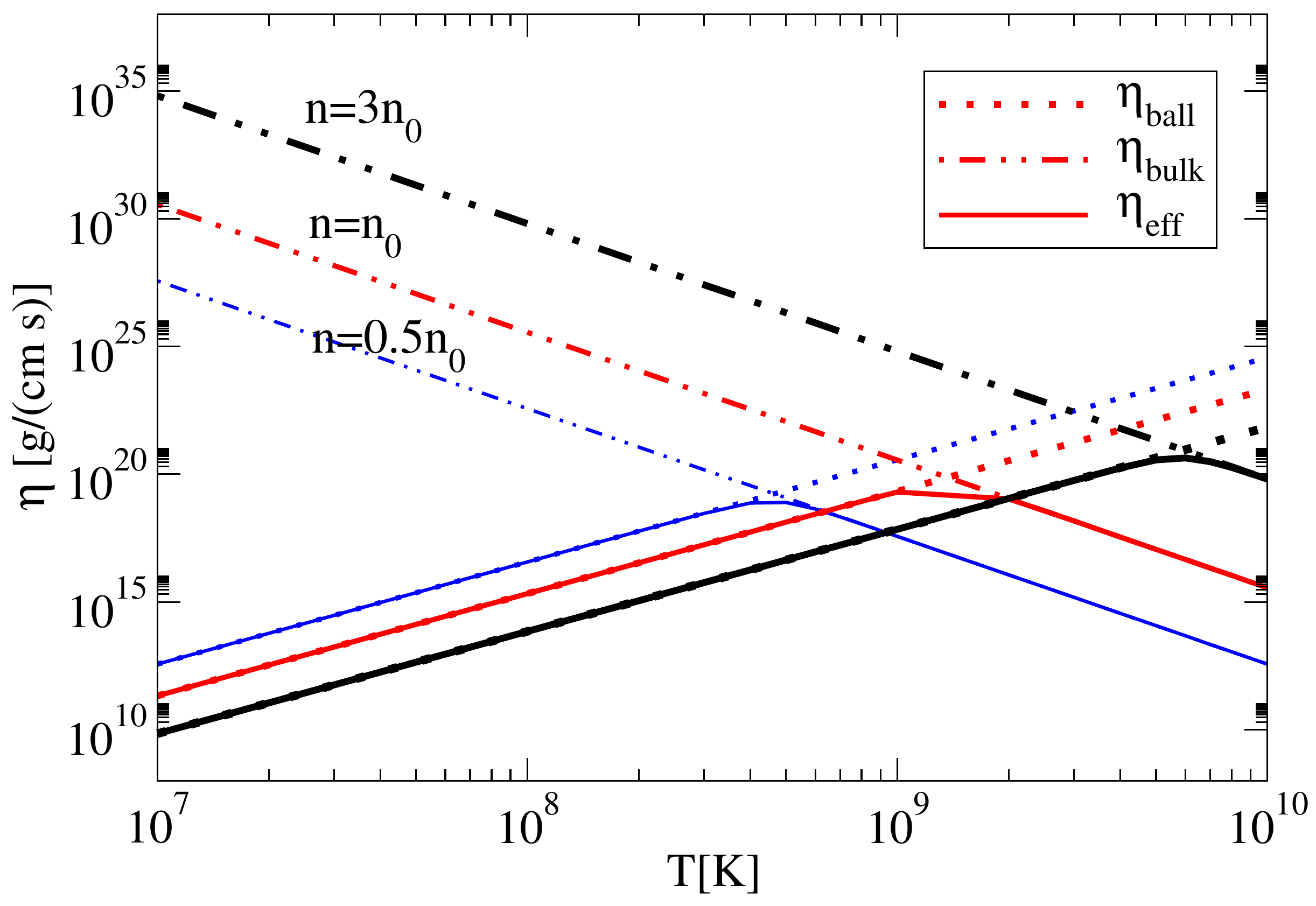}
\caption{Phonon contributions to the shear viscosity as a function of temperature for different particle densities in terms of the saturation density $n_0$. We show the ballistic viscosity 
$\eta_{\rm ball}$, the hydrodynamic viscosity $\eta_{\rm bulk}$, and the effective description $\eta_{\rm eff}$. 
{Figure adapted} from~\cite{Manuel:2012rd}.}
\label{fig:ph}
\end{figure}

We show the values of the ballistic $\eta_{\rm ball}$, the hydrodynamic $\eta_{\rm bulk}$, and the effective $\eta_{\rm eff}$ shear viscosities in Figure~\ref{fig:ph}. These contributions are displayed as a function of temperature for different particle densities in terms of the normal saturation density. These values for the density are expected to be found in the interior of neutron stars. For the ballistic contribution, we use the value of the core radius $R_c =9.96$ km, which corresponds to a density of 0.5 $n_0$ for a star of 1.93 ${\rm M_{sol}}$, as used in the next section.

\textls[-25]{We find three different regions that appear depending on the dominant contribution to the shear viscosity. Whereas for $T < 10^{8}~{\rm K}$, the shear viscosity is described by the ballistic contribution for all densities in the core, for larger temperatures, the transition from the ballistic to the hydrodynamic domain takes place, this intermediate regime being strongly dependent on the density. For $n \sim n_0$, this transition domain appears for $T\sim 10^{9}~{\rm K}$, while for larger temperatures, the hydrodynamic regime dominates. 
We also find that for a given value of the temperature, the ballistic viscosity decreases as the particle density increases, whereas the hydrodynamic viscosity exhibits the opposite behavior. Moreover, note that this present study and our previous analysis of the mean free path in Figure~\ref{fig:shear} (lower panel) are in good agreement.}

\section{Bulk Viscosities}
\label{sec:bulk}

Apart from the shear viscosity, another dissipative term that appears in the energy-momentum tensor is the bulk viscosity ($\zeta$). However, in superfluid matter, there exist four bulk viscosity coefficients~\cite{Khalatnikov:106134}.~Three of them, $\zeta_1, \zeta_3, \zeta_4$, are associated with dissipative processes, leading to entropy production related to the space--time-dependent relative motion between the superfluid and normal fluid components, whereas $\zeta_2$ is the analogue to the one in a normal superfluid.

Following the dynamical evolution of the phonon number density, developed by Khalatnikov~\cite{Khalatnikov:106134}, the bulk viscosities can be determined. This method is equivalent 
to calculating the bulk viscosities by means of the Boltzmann equation for phonons in the relaxation time approximation, as shown in~\cite{Escobedo:2009bh}.~In the case of small departures from equilibrium and small values of the normal and superfluid velocities, one obtains~\cite{Khalatnikov:106134}:
\be
\zeta_i = \frac{T}{\Gamma_{ph}} \, C_i \ , \qquad i=1,2,3,4 \ ,
\label{bulkstatic}
\ee 
with $\Gamma_{ph}$ the phonon decay rate and: 
\be
C_1 = C_4 = -I_1 I_2 \ , \qquad C_2 = I_2^2 \ , \qquad C_3 = I_1^2 \ ,
\label{coef}
\ee
where $I_1$ and $I_2$ are: 
\ba
I_1&=&\frac{60 T^5}{7c^7_s \pi^2}\left(\pi^2\zeta(3)-7\zeta(5)\right)\left(c_s\frac{\partial B}{\partial \rho}-B\frac{\partial c_s}{\partial \rho }\right)\,, \nonumber \\
I_2&=&-\frac{20 T^5}{7c^7_s \pi^2}\left(\pi^2\zeta(3)-7\zeta(5)\right)\left(2Bc_s+3\rho \left(c_s\frac{\partial B}{\partial \rho}-B\frac{\partial c_s}{\partial \rho }\right)\right)\,,
\label{i1-i2}
\ea
with $B=c_s \gamma$ and $\zeta(n)$ the Riemann zeta function. We should indicate that one has to consider the phonon dispersion law beyond linear order so as to have non-vanishing bulk viscosities. Moreover, due the Onsager symmetry principle, $\zeta_1=\zeta_4$, whereas positive entropy production leads to $\zeta_2, \zeta_3 \ge 0$ and $\zeta_1^2 \le \zeta_2 \zeta_3$

\begin{figure}[H]
{\includegraphics[width=12cm]{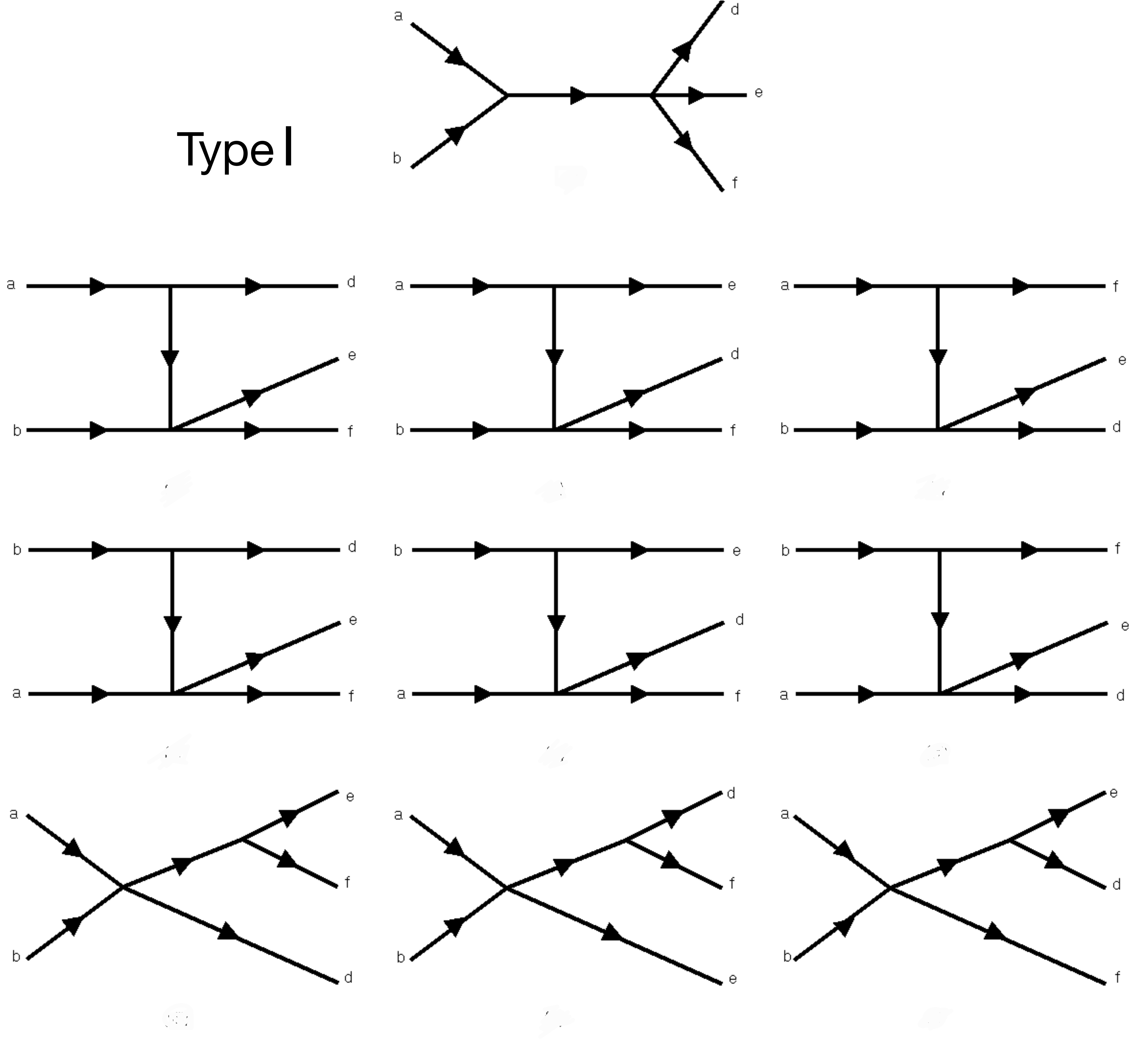}}
\caption{{Type I} diagrams are formed by one three-phonon vertex and one four-phonon vertex of $\mathcal{L}_{LO}.$ Figure {adapted from}~\cite{Manuel:2013bwa}. }
\label{type1}  
\end{figure}

In the case of astrophysical phenomena, the bulk viscosities are calculated for periodic perturbations, leading the system away from equilibrium. In~\cite{Bierkandt:2011zp}, it was shown that:
\be
\label{w-bulks}
\zeta_i (\omega) =\frac{1}{1+\left(\omega I^2_1 \,\frac{\partial \rho}{\partial n}\frac{\partial \rho}
{\partial \mu}\frac{T}{\Gamma_{ph}}\right)^2} \frac{T}{\Gamma_{ph}} \, C_i 
 \ , \qquad i=1,2,3,4 \ ,
\ee
defining a characteristic value for the frequency $\omega_c$ for the phonon collisions as: 
\be
\omega_c = \frac{1}{I^2_1 \,\frac{\partial \rho}{\partial n}
\frac{\partial \rho} {\partial \mu}}
\frac{\Gamma_{ph}}{T} \ .
\label{char}
\ee
Note that when $\omega \ll \omega_c$, one recovers the static bulk viscosity coefficients of Equation~(\ref{bulkstatic}).

\begin{figure}[H]
{\includegraphics[width=12cm]{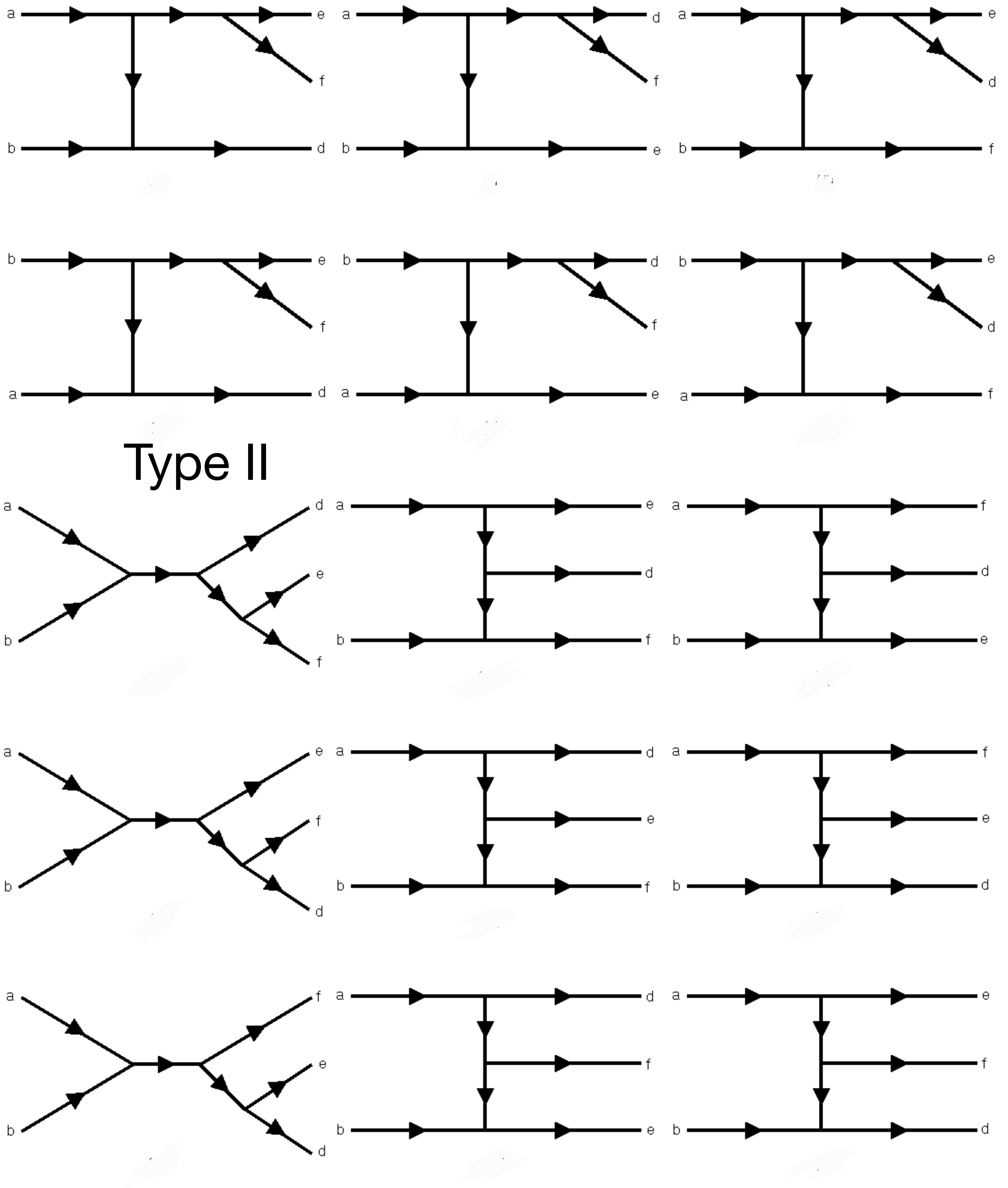}}
\caption{{Type II} diagrams are formed by three three-phonon vertices of $\mathcal{L}_{LO}$. Figure {adapted from}~\cite{Manuel:2013bwa}. }
\label{type2} 
\end{figure}

In order to go back to equilibrium after an expansion of the superfluid, the system needs to change the number of phonons. Therefore, $\Gamma_{ph}$ includes the first number-changing mechanisms allowed by kinematics. In the case of phonons with a negative $\gamma$ in the dispersion law, those are $2 \leftrightarrow 3$ collisions. The different $2 \leftrightarrow 3$ processes are given in \mbox{Figures~\ref{type1} and \ref{type2}} and can also be found in~\cite{Manuel:2013bwa}. Those are indicated as Type I, given in \mbox{Figure~\ref{type1}}, and Type II, given in Figure~\ref{type2}, using the phonon propagators with an NLO dispersion law, as discussed~\cite{Manuel:2013bwa}.

\begin{figure}[H]
\includegraphics[width=0.6\textwidth]{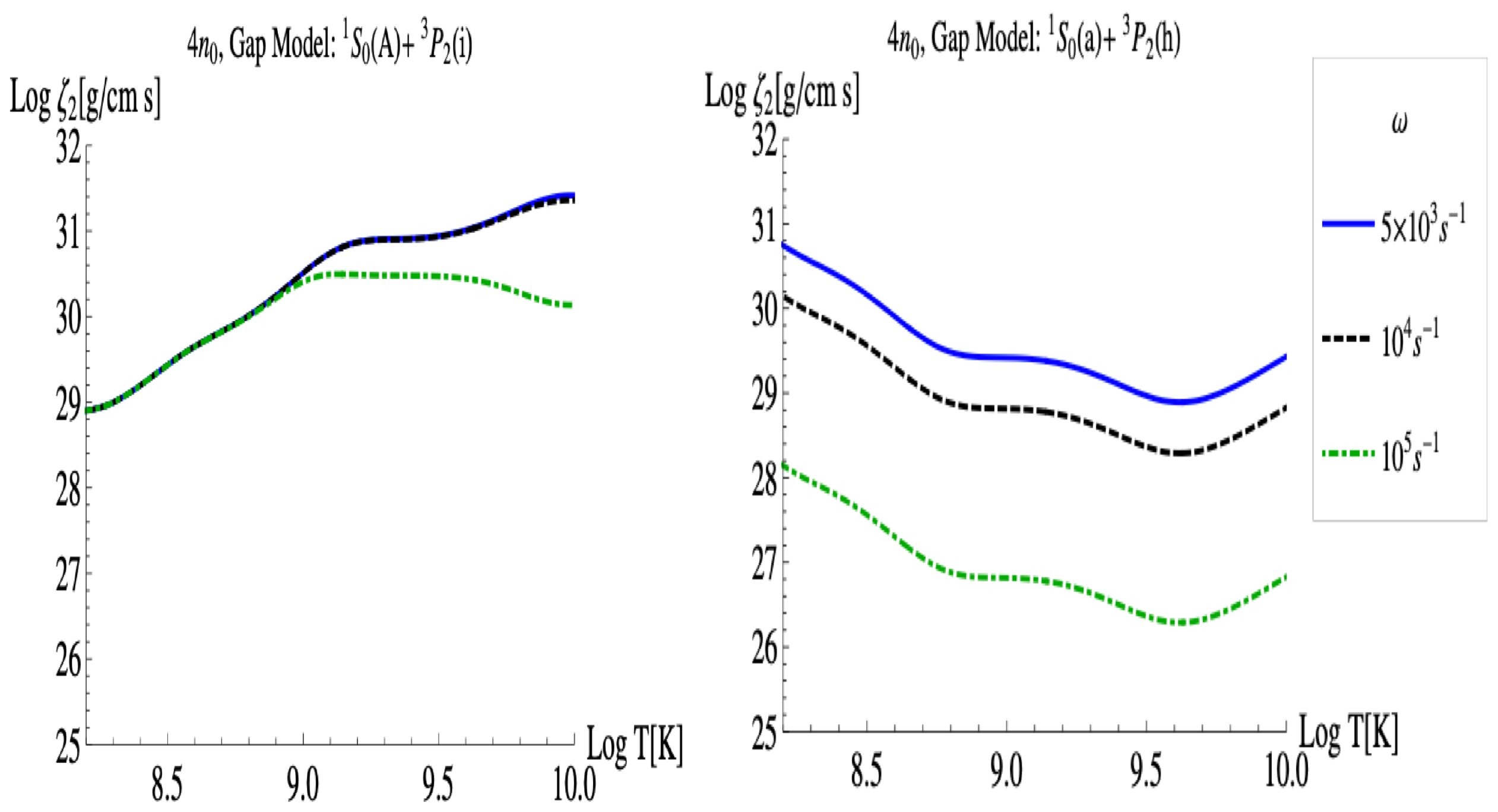} 
\caption{$\zeta_2$ {frequency-dependent} bulk viscosity coefficient as a function of the temperature for $4n_0$ and frequencies between $10^{3}$ and $10^{5}$ s$^{-1}$ for the two neutron gap models. Figure adapted from~\cite{Manuel:2013bwa,Tolos:2014wla}. Adapted from \cite{Tolos:2014wla} with the permission of AIP {Publishing}.} 
\label{fig:zeta2}
\end{figure}

The frequency-dependent $\zeta_2$ coefficient at $4n_0$ for frequencies of $\omega=10^3-10^5$~s$^{-1}$, typical for stellar pulsations, is shown as a function of temperature in Figure~\ref{fig:zeta2}. On the l.h.s. of Figure~\ref{fig:zeta2}, we show the $\zeta_2$ coefficient when using the $^1S_0(A)+^3P_2(i)$ neutron gap model. The value of $\zeta_2$ for $\omega=10^3$~s$^{-1}$ differs by more than 10$\%$ from its static value for only $T\gtrsim10^{10}$~K, whereas the difference is larger for $T\gtrsim10^9$~K in the case of $\omega= 10^5$~s$^{-1}$. As for the case when using the $^1S_0(a)+^3P_2(h)$ model on the r.h.s. of Figure~\ref{fig:zeta2}, we see that $\zeta_2$ strongly depends on the frequency. We did not plot the $\zeta_1$ and $\zeta_3$ coefficients, as we expect a similar behavior with frequency.

The frequency-dependent $\zeta_2$ bulk viscosity coefficient resulting from the collisions of superfluid phonons has to be compared to contributions coming from direct Urca~\cite{Haensel:2000vz} and modified Urca~\cite{Haensel:2001mw} processes. We see that phonon collisions are the leading contribution to the bulk viscosities in the core for $T\sim 10^9$~K and for typical radial pulsations. The different conclusion reached in~\cite{Manuel:2013bwa} is due to the fact that the comparison was performed only for 
Urca processes in normal matter, as the phonon collisions are the most important contribution until the opening of the Urca processes.
\section{The R-Mode Instability Window}
\label{sec:r-modes}

R-mode oscillations (modes for which the restoring force is the Coriolis force) of neutron stars have been extensively studied because they appear to be subject to the Chandrasekhar--Friedman--Schutz gravitational radiation instability in realistic astrophysical conditions (see, for example, reviews~\cite{Andersson:2000mf,Lindblom:2000jw}).~A rapidly rotating neutron star could emit a significant fraction of its rotational energy and angular momentum as gravitational waves, which can be detected by interferometers. If the timescales due to viscous processes in neutron star matter are shorter than the gravitational radiation driving timescale, r-modes are damped. 
There is a typical instability region at relatively high frequencies, whereas the star is stable for low frequencies, or at very low or high temperatures, due to viscous damping mechanisms~\cite{Lindblom:1998wf, Andersson:1998ze}. Given that the spin rate of various neutron stars falls in the instability window, the r-mode studies are trying to find new solutions to this~puzzle.

The r-mode instability window is determined by the computation of the different time scales associated with gravitational wave emission $\tau_{\rm GR}$ and with the different dissipative processes that could damp the r-mode. In this paper, we review the effect of the shear viscosity, $\tau_{\eta}$, in the r-mode instability window. The analysis of time scales involving the bulk viscosities due to superfluid phonons are more involved, as not only one, but three independent bulk viscosities appear, and these are a matter of future work. To find the instability window, we equate the time scales for gravitational wave emission and shear~viscosity,
\be
- \frac{1} { |\tau_{\rm GR} (\Omega) |} +\frac{1}{\tau_{\eta} (T)} = 0 \ ,
\ee
where the different time scales are given by:
 \be
\frac{1} { |\tau_{\rm GR} (\Omega) |} = \frac{32 \,\pi \,G\, \Omega^{2l+2}}{c^{2l+3}} \frac{(l-1)^{2l}}{\left( (2l+1)!! \right)^2} \left( \frac{l+2}{l+1} \right)^{2 l+2} \int^R_0 \rho r^{2 l+2} dr \ ,
\ee 
and: 
\be
\frac{1}{\tau_{\eta} (T)} = (l-1) (2l +1) \int^R_{R_c} \eta r^{2l} dr \left(\int^R_0 \rho r^{2l+2} dr \right)^{-1} \ ,
\label{ec-hidro-shear}
\ee
with the typical time scales summarized in~\cite{Lindblom:1998wf} (in C.G.S.
 units). 

In this paper, we concentrate our discussion on the study of the dominant r-modes with $l = 2$~\cite{Andersson:2000mf,Lindblom:1998wf} and the dominant damping mechanisms in the core of neutron stars. In fact, the shear viscosity coming from electron collisions is one of the most efficient damping mechanisms in neutron stars~\cite{Shternin:2008es}.~In the normal phase, the electron contribution to the shear viscosity has been found to be higher than the nucleon contribution. However, we note that there are some computations~\cite{Benhar:2009nr,Zhang:2010jf} that indicate that the nucleon contribution can be increased due to many-body effects and three-nucleon forces, so that it overcomes the electron contribution to the shear viscosity. Given that the electron collisions dominate over the nucleon collisions in the normal phase, one then expects that the same happens in the superfluid phase in the star. Therefore, we only present the contribution of the superfluid phonons to the shear viscosity, and we do not consider the contribution of the fermionic Bogoliubov quasiparticles, which, as far as we know, have not been computed so~far. 

With regards to the characteristic time associated with the phonon shear viscosity, we only consider the phonon contribution in the hydrodynamic regime. We then introduce a temperature-dependent lower cut ($R_c$) in the integral for the shear viscosity in \mbox{Equation~(\ref{ec-hidro-shear})}. The lower the limit is, the higher the density is for which phonons are hydrodynamic. The consideration of only phonon processes within the core comes from the fact that this characteristic time is obtained from considering dissipations in small volume elements throughout the star, and it is not related to the viscous process coming from the interaction of phonons with the crust in the ballistic regime. In any case, the dissipation associated with electron collisions takes over whenever the phonons are not in the hydrodynamic~regime.

\begin{figure}[H]
\includegraphics[width=0.65\textwidth]{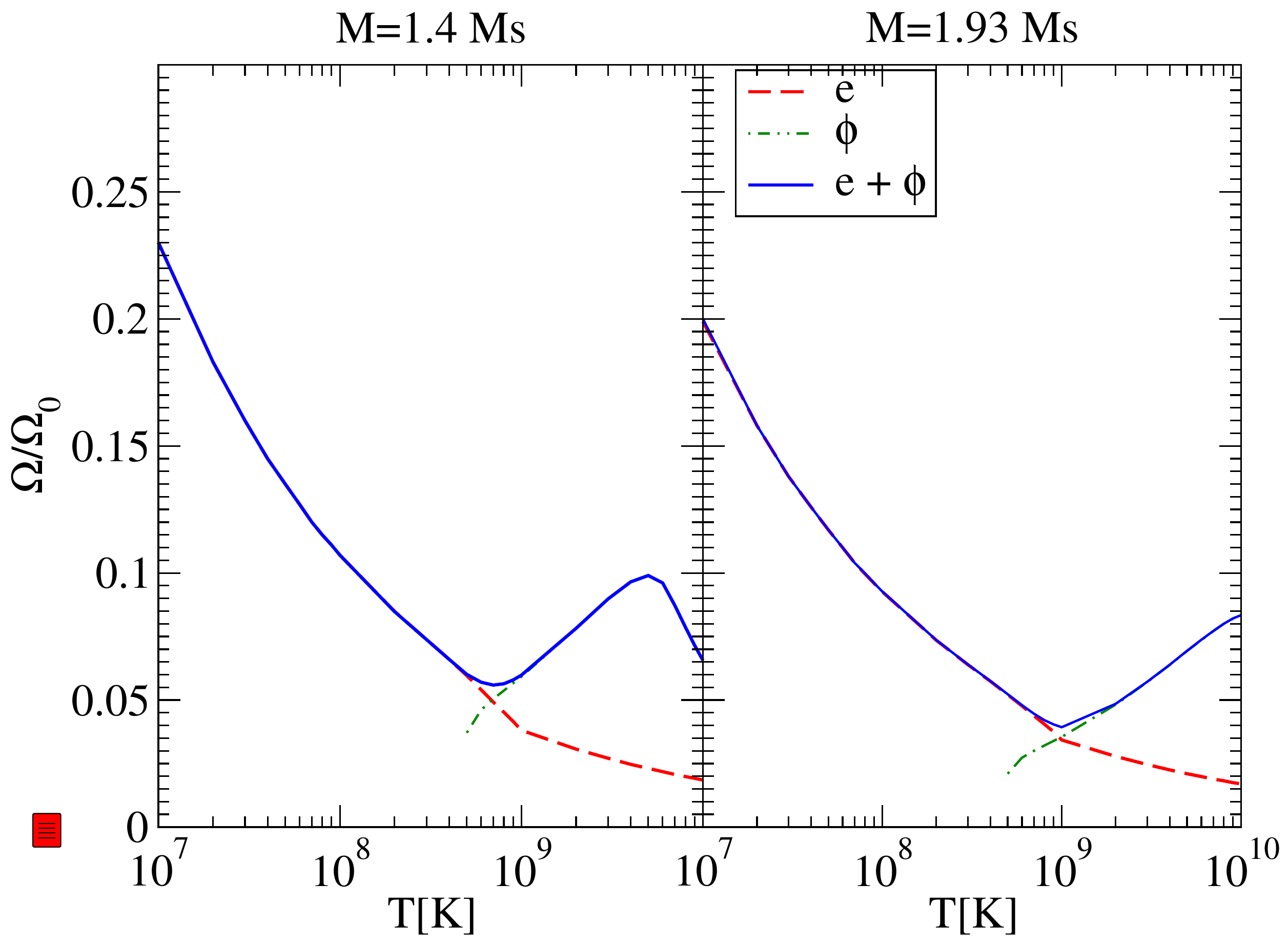}
\caption{R-mode instability window for superfluid neutron stars. The Kepler
(normalized) frequency is displayed for two neutron star masses (1.4 ${\rm M_{s}}$ and 1.93 ${\rm M_{s}}$) as a function of temperature for the electron, phonon, and electron+phonon dissipative processes. Here, $\Omega_0 =\sqrt{G \pi \bar{\rho}}$, where $\bar{\rho}$ is the average mass density of the star. Figure adapted from~\cite{Manuel:2012rd,Tolos:2014wla}. Adapted from \cite{Tolos:2014wla} with the permission of { AIP Publishing}.}
\label{fig:freq}
\end{figure}

The r-mode instability window coming from the different shear viscous damping mechanisms is shown in Figure~\ref{fig:freq}. The results are displayed for 1.4 ${\rm M_{sun}}$ and 1.93 ${\rm M_{sun}}$ mass configurations. The quantity $\Omega_0 =\sqrt{G \pi \bar{\rho}}$ is the Kepler frequency, given in terms of the average mass density of the star, $\bar{\rho}$. For the case of electrons, we show the estimate coming from the longitudinal and transverse plasmon exchange in superconducting matter with a transition temperature of $T_{cp} \sim 10^9~{\rm K}$~\cite{Shternin:2008es} (dashed lines). As for the hydrodynamic phonons, we indicate our results in dashed-double-dotted lines. Finally, we add both contributions, indicated in solid lines. By analyzing those results, we find that the electron contribution dominates for low temperatures, whereas the contribution of hydrodynamic phonon dissipation is relevant for $T \gtrsim7 \times 10^8~{\rm K}$ for a star of 1.4 ${\rm M_{sun}}$ and $T \gtrsim10^9~{\rm K}$ for 1.93 ${\rm M_{sun}}$.

\section{Thermal Conductivity}
\label{sec:thermal}

Another transport coefficient that we would like to discuss is the thermal conductivity due to superfluid phonon collisions. The thermal conductivity $\kappa$
relates the heat flux to the temperature gradient in the hydrodynamic regime:
 \ba
{\bf q}=-\kappa \nabla T \ .
\label{defi}
 \ea
 
 In order to obtain this coefficient, we use variational methods to solve the transport equation, as done for the shear viscosity. In this case, the thermal conductivity is given by~\cite{Manuel:2014kqa}:
\ba
\kappa \geq \left(\frac{4a_1^2}{3 T^2}\right) A_1^2 M^{-1}_{11},
\label{kappa_var}
\ea
with $M^{-1}_{11}$ the (1,1) element of the inverse of a matrix of $N \times N$ dimensions, with $N$ being treated variationally. This matrix contains different elements that are multidimensional integrals with thermally weighted phonon scattering matrix. Note that, as in the case of bulk viscosities, the thermal conductivity requires the dispersion law at NLO, because the thermal conductivity vanishes with a linear dispersion law~\cite{Braby:2009dw}.

\begin{figure}[H]
\includegraphics[width=0.58\textwidth]{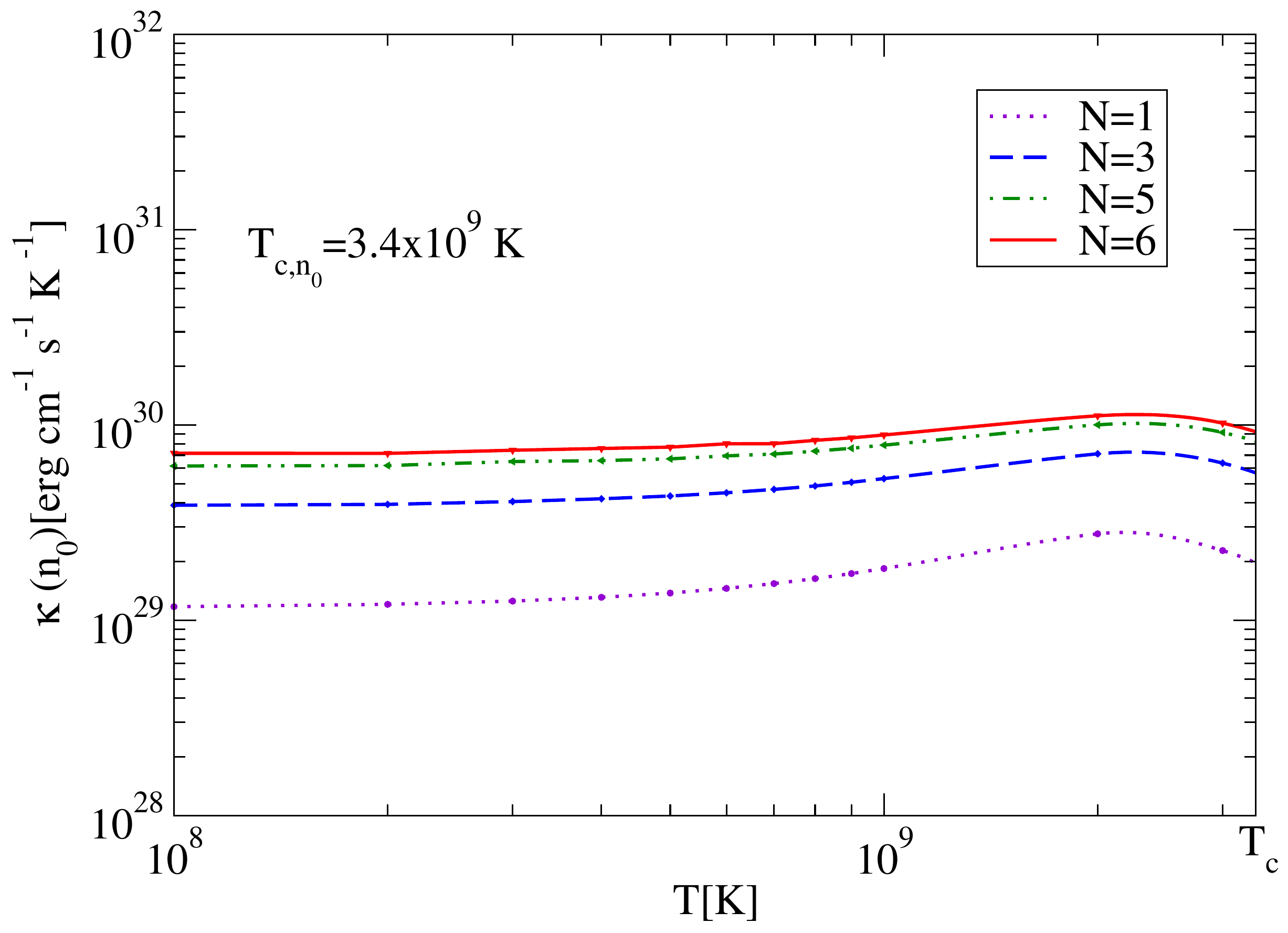}
\includegraphics[width=0.6\textwidth]{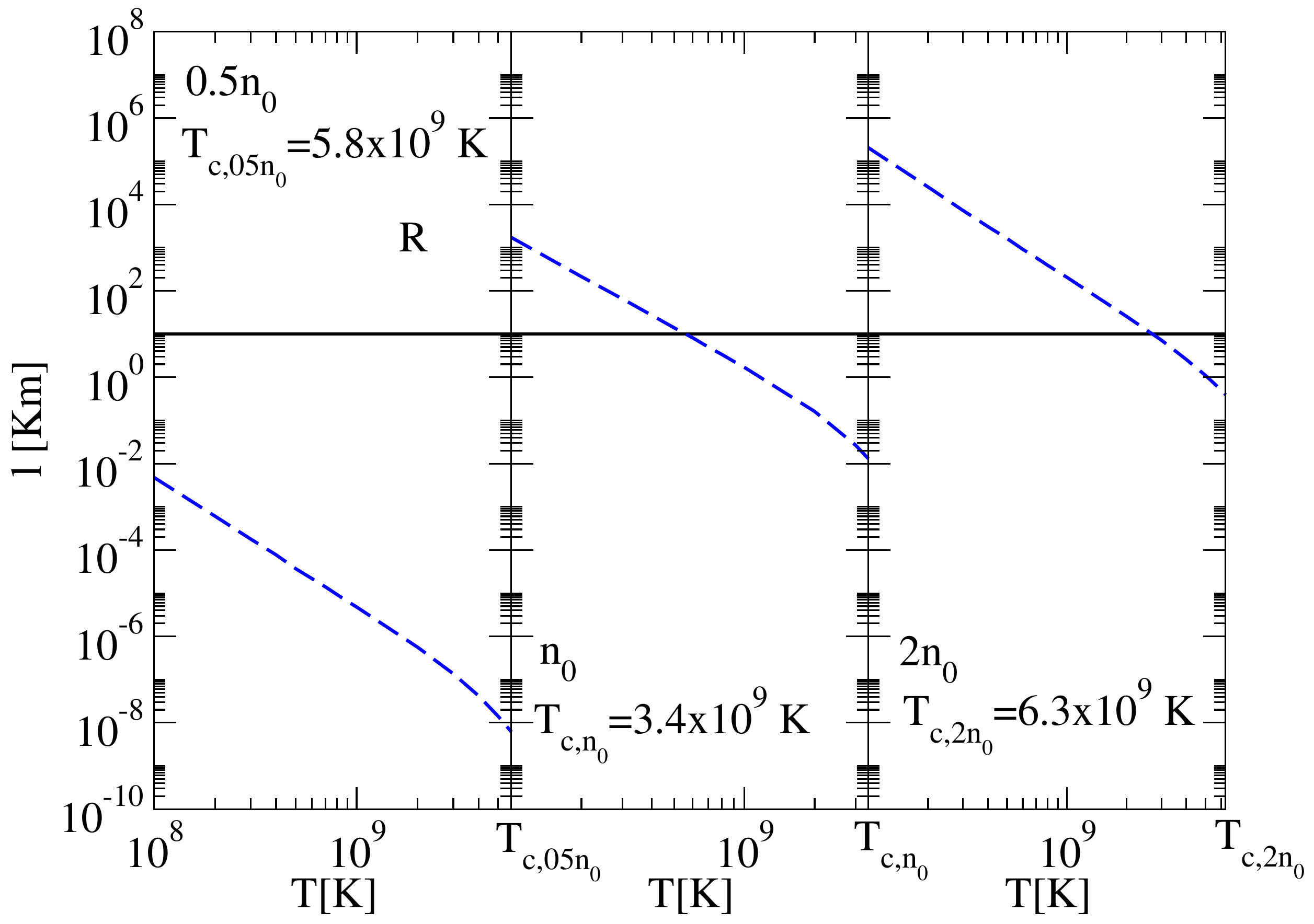}
 \caption{{\textbf{Upper  panel}}: Thermal conductivity due to superfluid phonons by means of a variational calculation up to order $N=6$ for $n_0$ as a function of temperature. We use the $^1S_0(A)$$+$$^3P_2(i)$ gap model. \textbf{Lower panel}: Mean free path obtained from the phonon thermal conductivity using the $^1S_0(A)$$+$$^3P_2(i)$ model for the gap as a function of temperature for three different densities. The critical temperature shown is $T_c=0.57 \Delta$, whereas we compare the mean free path to the radius of the star ($R=10$ km). Figures adapted from~\cite{Manuel:2014kqa, Tolos:2014wla}. Adapted from \cite{Tolos:2014wla} with the permission of { AIP Publishing}.} 
 \label{fig:var-mfp}
\end{figure}

We show the variational calculation of the thermal conductivity up to order $N=6$ for nuclear saturation density $n_0$ as a function of temperature in the upper panel of Figure~\ref{fig:var-mfp}. The last value $N$ is determined once the final results do not deviate more than $10\%$ from those in the previous iteration. The value $T_c$ is the end temperature, which turns out to be $T_c=0.57 \Delta(n_0)=3.4 \times 10^9$ K, with $\Delta(n)$ from Figure~\ref{speeds-gap} for the $^1S_0(A)$$+$$^3P_2(i)$ gap model.
We find that for T$ \lesssim 10^9$ K, well below $T_c$, the thermal conductivity scales as $\kappa \propto 1/\Delta^6$. The proportionality factor depends on the EoS. This temperature independent behavior was also observed for the color-flavor locked superfluid~\cite{Braby:2009dw}. Note that close to $T_c$, higher order corrections in energy and momentum might be expected in the phonon dispersion law and self-interactions.

Associated with the thermal conductivity, one can also determine the mfp for phonons, which is different from the one coming from the shear viscosity (see \mbox{Equations~(\ref{eq1}) and (\ref{eq2})} to compare). The mfp resulting from the thermal conductivity is given by:
\ba
l= \frac{\kappa}{\frac{1}{3} c_v c_s} ,
\ea
\ba
c_v= \frac{2 \pi^2}{15 c_s^3} \left( T^3 + \frac{25 \gamma}{ 7} \frac{(2 \pi)^2}{c_s^2} T^5 \right) \ ,
\ea
with $c_v$ the heat capacity for phonons~\cite{Khalatnikov:106134}.

In the lower panel of Figure~\ref{fig:var-mfp}, we display the mfp of phonons in $\beta$-stable neutron star matter for different densities ($0.5 n_0$ in the left panel, $n_0$ in the middle panel, and $2 n_0$ in the right panel) as a function of the temperature in the case of the $^1S_0(A)$$+$$^3P_2(i)$ gap model. For comparison, we also show the radius of the star of 10 km with a horizontal line. We find that for $n= 0.5 n_0$, the superfluid phonon mfp stays below the radius of the star. This is also the case for 
$n= n_0$ and T $\gtrsim6\times10^8$ K and for 2$n_0$ and T $\gtrsim3\times10^9$ K. Moreover, $l \propto 1/T^3$, coming from the temperature-independent behavior of the thermal conductivity. For smaller values of the gap, as the $^1S_0(a)$$+$$^3P_2(h)$ case, the thermal conductivity would be orders of magnitude higher than the previous case, being away from the hydrodynamic regime. One could then define a ballistic thermal conductivity and carry out a similar approach as that of the shear viscosity in Section~\ref{ball-sec}.

Again, our results must be compared to those coming from electrons and muons in neutron stars. The electron-muon contribution to the thermal conductivity was analyzed in~\cite{Shternin:2007ee}.~Compared to these results, we find that phonons in the hydrodynamic regime dominate the thermal conductivity in neutron stars~\cite{Manuel:2014kqa}. We also conclude that if the contribution of electrons-muons and phonons to the thermal conductivity become comparable, electron-phonon collisions could play an important role. Simple estimates were performed in~\cite{Bedaque:2013fja}. This topic deserves further studies.

\section{Neutrino Emissivity and the Superfluid Phonon}
\label{sec:neutrino}

The cooling of a neutron star is very much affected 
if superfluidity is achieved in its core. Close to the superfluid phase transition, the neutrino emissivity is
dominated by the formation/breaking of Cooper pairs~\cite{Flowers:1976ux,Voskresensky:1987hm,Yakovlev:1998wr}.
However, 
at much lower temperatures, these processes are exponentially suppressed, and the neutrino emissivity is dominated by scatterings involving the Goldstone modes of the system, as these couple with the $Z^0$ 
electroweak gauge boson. In~\cite{Bedaque:2003wj,Bedaque:2013rya}, the neutrino emissivity involving the angulons, the Goldstone modes associated with the spontaneous breaking of the rotational symmetry that occur in a $^3P_2$ neutron superfluid 
phase, were considered, while the contribution associated with the superfluid phonons was not taken into account. Later results on pair breaking emissivity were given in~\cite{Kolomeitsev:2008mc}.


The superfluid phonon interacts with the electroweak $Z^0$ gauge boson, which, in turn, can decay into a neutrino-antineutrino pair. Standard EFT techniques can be used to determine the Lagrangian associated with these electroweak processes
~\cite{Bedaque:2003wj}:
\be
{\cal L}_{\rm EW} = - f_0 C_V Z^0_0 \partial_0 \phi + \cdots + g_{Z{\bar \nu} \nu} Z_\mu^0 {\bar \nu} \gamma^\mu (1 - \gamma_5) \nu \ ,
\ee
where: 
\be
C_V^2 = \frac{ G_F M^2_Z}{2 \sqrt{2}} \ , \qquad
g^2_{Z{\bar \nu} \nu} = \frac{ G_F M^2_Z}{2 \sqrt{2}} \ ,
\ee
where $G_F$ is the Fermi constant and $M_Z$ is the mass of the $ Z_\mu$ electroweak gauge boson,
and
\be
f_0 = 2 \frac{\sqrt{\rho}}{c_s m} \ ,
\ee
where $\rho$ is the mass density and $m$ is the neutron mass.
Note that in~\cite{Bedaque:2003wj}, $f_0$ was the superfluid decay constant, different from the constant used here. The difference comes in how
the superfluid field is normalized with respect to the phase of the neutron-neutron condensate, which in our case is not the same as that used in~\cite{Bedaque:2003wj}; see Equation~(\ref{normalization}).

Energy and momentum conservation prevent the possibility that a superfluid phonon may decay into a neutrino-antineutrino pair. However, the process:
\be
\label{scatteringprocess}
\phi + \phi \rightarrow \phi +\nu + {\bar \nu} 
\ee
is kinematically allowed. This process has been considered in the color-flavor locked superfluid quark matter~\cite{Jaikumar:2002vg}.
A rough estimate of the neutrino emissivity $\cal E$ associated with this process reveals that it is very much suppressed, involving the $\phi \phi$ binary scattering, and
with the dependence of the four-phonon coupling constant. The scattering amplitude of $\phi \phi$ collisions goes as $\sim \lambda p^4$, where $p$ is the
phonon momenta, which should be of the order of the temperature $T$. Thus, parametrically, a naive dimensional analysis gives:
\be
{\cal E} \sim G_F^2 T^9 \frac{T^8}{m^4 \rho} \ .
\label{estimate-SPHnu}
\ee

\textls[-15]{In the temperature regime where the Goldstone modes might be relevant (say \mbox{$T < 10^9 K$})}, the ratio in Equation~(\ref{estimate-SPHnu}) suppresses
very much the strength of this channel, in front of that involving only the angulons, which parametrically behave as ${\cal E} \sim G_F^2 T^9$~\cite{Bedaque:2003wj}, for example. Superfluid phonons do not thus play any relevant role in the cooling of the star by neutrino emission.

\section{Superfluid Phonons in the Presence of Gravity and in a Moving Background}
\label{sec:gravity}

In the computation of the transport coefficients in superfluid neutron stars associated with superfluid phonons, we assumed that these move in
a static medium, and we ignored the effect of the gravitational field. These two effects might be easily incorporated into our EFT approach.
Let us explain how.

In what follows, we deal with the case of a relativistic superfluid, where all the EFT techniques used for the non-relativistic case also apply,
with minor changes. In particular, we use $X = (g^{\mu \nu} D_\mu \varphi D_\mu \varphi)^{1/2}$, with $D_\mu \varphi = \pa_\mu \varphi - \mu_r A_\mu$, and $A_\mu = (1,0,0,0)$, and $g^{\mu \nu}$ is the gravitational metric. The
 non-relativistic case can be obtained from the relativistic case~\cite{Nicolis:2017eqo}, by simply approximating the $X$ function, taking into account also that a non-relativistic
potential is defined as $\mu = \mu_r - m $ from the relativistic one $\mu_r$. In this case, it is only necessary to keep the $(0,0)$ component of the metric, which is related to
the Newtonian gravitational potential by $\delta g_{00} = - 2 \Phi$. Thus, in the presence of a gravitational potential, Equation~(\ref{LO-Lagran}) is replaced by:
\be
X_{\rm g} = \mu-\partial_t\varphi-\frac{({\bf \nabla}\varphi)^2}{2m} - m \Phi \ .
\ee
Thus, even if the superfluid phonons are massless, they are affected by the presence of a gravitational field, even in the non-relativistic case.

Even if we ignore the effects of a gravitational metric, as we will do in what follows, taking into account the effects of a moving superfluid medium in the superfluid phonons
is much more conveniently done with the use a gravitational analogue model~\cite{Manuel:2007pz,Mannarelli:2008jq,Mannarelli:2020ebs}. This was first suggested in~\cite{Volovik:2000ua}, but we will follow a different approach here, using the superfluid phonon EFT.
From the expression of the phonon Lagrangian in terms of $X$, it is possible to derive the EFT of the phonons moving in the background of the superfluid. The superfluid phonon is the Goldstone boson associated with the breaking of the $U(1)$ symmetry, and it
can be introduced as the phase of the quantum condensate. However, according to Landau's two-fluid model of superfluidity, the gradient of the phase of the condensate defines
the velocity of the pure superfluid component. Then, it should be possible to decompose the 
phase of the condensate into two fields, the first describing the hydrodynamic variable and the second describing the quantum fluctuations associated
with the phonons. 
Thus, we write:
\begin{equation}
\label{split}
\varphi (x) = \bar \varphi(x) + \phi(x) \ .
\end{equation}
This splitting implies a separation of scales: the background field $\bar \varphi(x)$
is associated with the long-distance and long-time scales, while the fluctuation $\phi(x)$ is
associated with rapid and small-scale variations and is identified with the superfluid phonon. The gradient 
of $\bar \varphi(x)$ is proportional to the hydrodynamic velocity:
\begin{equation} \label{svelocity}
 v_\rho = - \frac{D_\rho \bar\varphi}{\bar \mu} \ , \qquad \bar \mu \equiv (D_\rho \bar \varphi D^\rho \bar \varphi)^{1/2} .
\end{equation}

The classical equations of motion associated with $\bar \varphi(x)$ can be conveniently expressed as the hydrodynamic equations
of a perfect relativistic fluid:

\begin{equation}
\label{S-hy-1}
 \partial_\nu (\tilde n_0 v^\nu) = 0 \ ,
\end{equation}
 where: \be \tilde n_0 =\frac{dP}{d \mu} \Big |_{\mu =\bar \mu} \ee is interpreted
as the particle density.
The energy-momentum tensor can be written in terms of the velocity
defined in Equation~(\ref{svelocity}) and Noether's energy-density $\rho_0$,
 \begin{equation}
\label{S-hy-2}
T^{\rho \sigma}_0 = (\tilde n_0 \bar \mu) v^\rho v^\sigma - \eta^{\rho \sigma} P_0 = (\rho_0 + P_0) v^\rho v^\sigma - \eta^{\rho \sigma} P_0 \ ,
 \end{equation}
where we write $\rho_0 + P_0 = \tilde n_0 \bar \mu$, with $P_0$ the pressure evaluated at
$\bar \mu$.
 The energy-momentum tensor is conserved:
\begin{equation} \label{Tconserved}
\partial_\rho T^{\rho \sigma}_0 = 0\ ,
\end{equation}
and traceless $ T^\rho_{0 \,\rho}=0$.

From the low energy effective action of the system:
\begin{equation}
S[\varphi] = \int d^4 x \, {\cal L}_{\rm eff}[\partial \varphi] \,,
\end{equation}
we deduce the effective action for the phonon field expanding around the stationary point corresponding to the classical solution $\bar \varphi$:
\ba \label{action2}
S[\varphi] &=& S[\bar \varphi] + \frac 12 \int d^4 x \,\frac{ \partial^2 {\cal L}_{\rm eff} }
{\partial(\partial_\mu\varphi) \partial(\partial_\nu \varphi)} \Bigg \vert_{\bar \varphi}\partial_\mu \phi\, \partial_\nu \phi + \cdots \,, \\
& = & = \frac 12 \int d^4 x \sqrt{- {\cal G} } \, {\cal G}^{\mu \nu} \partial_\mu \phi\, \partial_\nu \phi \ , 
\ea
where we define the acoustic metric tensor:
\begin{equation}
\label{phonon-metric}
{\cal G}^{\mu \nu} = 
\eta^{\mu\nu} + \left(\frac {1}{c_s^2} - 1 \right) v^\mu v^\nu \ ,
\end{equation}
and the determinant ${\cal G} = 1/{\rm det} | {\cal G}^{\mu \nu}|$. Thus, in the background of a moving superfluid, the superfluid phonon propagates as in the background of the so-called 
acoustic metric. Note that in the presence of a real gravitational field, one simply has to write the corresponding metric $g^{\mu \nu}$ above instead
of the Minkowskian metric $\eta^{\mu \nu}$.

The transport equation associated with the superfluid phonons should then incorporate the effect of the acoustic metric. It should be written as:
\begin{equation}
\label{Blotzmman}
L[f] \equiv p^\alpha \frac{\partial f}{\partial x^\alpha} - \Gamma^\alpha_{\beta\gamma} p^\beta p^\gamma \frac{\partial f}{\partial
p^\alpha} = C[f]
\end{equation}
that is the general relativistic version of the Boltzmann equation. The Christoffel symbols of the phonon transport equation are those
related to the acoustic metric. 
Quite interestingly, from this approach, it is possible to deduce that even in the collisionless limit, 
the number of phonons might not be conserved, as they are covariantly conserved:
\begin{equation}
 \partial_\nu n_{\rm ph}^\nu + \Gamma^\mu_{\mu \nu} n_{\rm ph}^\nu = 0\; .
\end{equation}
This means that even in collisionless processes, the number of phonons might not be conserved.
 This is reflected in the second term in the l.h.s of the above equation where a Christoffel symbol appears, signaling that the propagation of phonons is taking place in ``curved'' space-time.
Therefore, we can write:
\begin{equation}
\Gamma^\mu_{\mu \alpha} = \frac{1}{\sqrt{- \cal G}} \partial_\alpha \sqrt{- \cal G}=
\frac{1}{c_s} \frac {\partial c_s}{\partial x^\alpha} \, .
\end{equation}
Thus, the variation of the speed of sound inside the star might act as a source, or sink, of superfluid phonons.

In all our developments presented in this review article, we ignored all the above effects to simplify the computations.
We might expect that if the mfp of the phonons is shorter than the variation of both the gravitational potential and
 of
the speed of sound, we can simply ignore the effects discussed in this Section.~Unfortunately, the phonon mfp tends to
increase when the temperature drops. Thus, the evaluation of transport should be reformulated along the lines discussed here.

An interesting observation was formulated in~\cite{Esposito:2018sdc}. While it was generally believed that sound waves do not transport mass, in that reference, it
was claimed that they do carry gravitational mass if non-linear order effects are considered. That is to say, they are affected by gravity, as we have just seen, and also generate a tiny gravitational field. In particular, this applies to the superfluid phonon field. The superfluid phonon can be considered as the quanta of sound waves, and even in the non-relativistic case, an associated wave packet with energy E carries a gravitational mass: 
\be
M \approx - \frac{\rho}{c_s} \frac{d c_s}{d \rho} \frac{E}{c_s} \ .
\ee
 It is however not obvious how this mass might have an effect in the
transport phenomena of neutron stars, as claimed in~\cite{Esposito:2018sdc}.

\section{Summary}
\label{sec:summary}

We present an overview of the computation of the shear and bulk viscosities together with the thermal conductivity due to superfluid phonons inside neutron stars, based on an effective field theory for the interaction among superfluid phonons.~The effective field theory approach is universal and valid for different superfluid systems.~As the superfluid phonons couple to the $Z$ electroweak gauge boson, they open a channel to
neutrino emissivity, but we checked that it is very much suppressed, so it can hardly affect the cooling of the star.

\textls[-15]{With regard to the shear viscosity, we found that the shear viscosity coming from binary collisions of superfluid phonons scales with $1/T^5$, as seen in $^4$He and cold Fermi gases at unitarity. Whether the temperature dependence is a universal feature, the evolution of the shear viscosity with density is determined by the equation of state under beta-stable~equilibrium.} 

\textls[-15]{As for the bulk viscosities in superfluid neutron matter, we see that the bulk viscosity coefficients are highly dependent on the superfluid neutron matter gap.~Nevertheless, phonon-phonon collisions rule the bulk viscosity over the Urca and modified Urca~processes.}

We further studied the r-mode instability in neutron stars and the consequences of the shear viscosity coming from superfluid phonons. We determined that the r-mode instability window would be modified for 
$T\gtrsim 10^8-10^9$~K, depending on the exact neutron star mass configuration.

Furthermore, we determined a temperature independent behavior of the thermal conductivity due to phonons well below the transition temperature, while scaling as
$1/\Delta^6$, similar to color-flavor locked phase. Furthermore, the thermal conductivity is dominated by phonon-phonon interactions in comparison with electron-muon collisions for densities in the core of neutron stars. 

We finally discuss how the superfluid phonon effective field theory, and ultimately their interactions, is modified in the presence of
a gravitational field, or by taking into account that the superfluid is not at rest. In particular, given that the mean free path of phonons tends to increase when the temperature drops, it would be interesting to evaluate the effect of gravity on the transport coefficients we reviewed.

\vspace{6pt} 



\authorcontributions{{The} authors have read and agreed to the published version of the manuscript.} 

\funding{This research received no external funding.}

\acknowledgments{We thank Jaume Tarr\'us and Sreemoyee Sarkar for past collaborations. This research was supported by the Spanish Ministerio de
Econom\'ia y Competitividad under Contract FPA2016-81114-P, Ministerio de Ciencia e Innovaci\'on under Contract PID2019-110165GB-I00, and by the EU STRONG-2020 project under the program H2020-INFRAIA-2018-1, Grant Agreement No. 824093. This work was also supported by the COST Action CA16214 PHAROS: The multimessenger physics 
 and astrophysics of neutron stars.}

\conflictsofinterest{The authors declare no conflict of interest.} 


\end{paracol}
\reftitle{References}






\begin{thebibliography}{999}

\bibitem[Landau(1941)]{Landau:1941vsj}
Landau, L.D.
\newblock {The theory of superfuidity of helium II}.
\newblock {\em J. Phys. (USSR)} {\bf 1941}, {\em 5},~71--100.

\bibitem[Migdal(1960)]{Migdal}
Migdal, A.{Superfluidity and the moments of inertia of nuclei.} 
\newblock {\em Sov. Phys.J. Exp. Theor. Phys.} {\bf 1960}, {\em 10},~176.

\bibitem[Sedrakian and Clark(2019)]{Sedrakian:2018ydt}
Sedrakian, A.; Clark, J.W.
\newblock {Superfluidity in nuclear systems and neutron stars}.
\newblock {\em Eur. Phys. J. A} {\bf 2019}, {\em 55},~167,
\newblock
  doi:{\changeurlcolor{black}\href{https://doi.org/10.1140/epja/i2019-12863-6}{\detokenize{10.1140/epja/i2019-12863-6}}}.

\bibitem[Page \em{et~al.}(2013)Page, Lattimer, Prakash, and
  Steiner]{Page:2013hxa}
Page, D.; Lattimer, J.M.; Prakash, M.; Steiner, A.W.
\newblock {Stellar Superfluids}. \emph{arXiv} {\bf 2013}, arXiv:1302.6626.

\bibitem[Page and Reddy(2012)]{Page:2012zt}
Page, D.; Reddy, S.
\newblock {Thermal and transport properties of the neutron star inner crust}. \emph{arXiv} 
  {\bf 2012}, arXiv:1201.5602.

\bibitem[Manuel and Tolos(2011)]{Manuel:2011ed}
Manuel, C.; Tolos, L.
\newblock {Shear viscosity due to phonons in superfluid neutron stars}.~{\em Phys. Rev. D} {\bf 2011}, {\em 84},~123007, doi:{\changeurlcolor{black}\href{https://doi.org/10.1103/PhysRevD.84.123007}{\detokenize{10.1103/PhysRevD.84.123007}}}.

\bibitem[Manuel and Tolos(2013)]{Manuel:2012rd}
Manuel, C.; Tolos, L.
\newblock {Shear viscosity and the r-mode instability window in superfluid
  neutron stars}.
\newblock {\em Phys. Rev. D} {\bf 2013}, {\em 88},~043001,
\newblock
  doi:{\changeurlcolor{black}\href{https://doi.org/10.1103/PhysRevD.88.043001}{\detokenize{10.1103/PhysRevD.88.043001}}}.

\bibitem[Manuel \em{et~al.}(2013)Manuel, Tarrus, and Tolos]{Manuel:2013bwa}
Manuel, C.; Tarrus, J.; Tolos, L.
\newblock {Bulk viscosity coefficients due to phonons in superfluid neutron
  stars}.
\newblock {\em J. Cosmos. Aastro. Phys.} {\bf 2013}, {\em 7},~3,
\newblock
  doi:{\changeurlcolor{black}\href{https://doi.org/10.1088/1475-7516/2013/07/003}{\detokenize{10.1088/1475-7516/2013/07/003}}}.

\bibitem[Manuel \em{et~al.}(2014)Manuel, Sarkar, and Tolos]{Manuel:2014kqa}
Manuel, C.; Sarkar, S.; Tolos, L.
\newblock {Thermal conductivity due to phonons in the core of superfluid
  neutron stars}.
\newblock {\em Phys. Rev. C} {\bf 2014}, {\em 90},~055803,
\newblock
  doi:{\changeurlcolor{black}\href{https://doi.org/10.1103/PhysRevC.90.055803}{\detokenize{10.1103/PhysRevC.90.055803}}}.

\bibitem[Son(2002)]{Son:2002zn}
Son, D.
\newblock {Low-energy quantum effective action for relativistic superfluids}. \emph{arXiv} 
  {\bf 2002}, hep-ph/0204199.

\bibitem[Son and Wingate(2006)]{Son:2005rv}
Son, D.; Wingate, M.
\newblock {General coordinate invariance and conformal invariance in
  nonrelativistic physics: Unitary Fermi gas}.
\newblock {\em Ann. Phys.} {\bf 2006}, {\em 321},~197--224,
\newblock
  doi:{\changeurlcolor{black}\href{https://doi.org/10.1016/j.aop.2005.11.001}{\detokenize{10.1016/j.aop.2005.11.001}}}.

\bibitem[Khalatnikov and Khalatnikov(1965)]{Khalatnikov:106134}
Khalatnikov, I.M.; Khalatnikov, I.M.
\newblock {\em {An Introduction to the Theory of Superfluidity}}; Frontiers in
  Physics, Benjamin: New York, NY,  USA, 1965.
\newblock (In Russian)

\bibitem[Escobedo and Manuel(2010)]{Escobedo:2010uv}
Escobedo, M.A.; Manuel, C.
\newblock {Effective field theory and dispersion law of the phonons of a
  non-relativistic superfluid}.
\newblock {\em Phys. Rev. A} {\bf 2010}, {\em 82},~023614,
\newblock
  doi:{\changeurlcolor{black}\href{https://doi.org/10.1103/PhysRevA.82.023614}{\detokenize{10.1103/PhysRevA.82.023614}}}.

\bibitem[Bedaque \em{et~al.}(2003)Bedaque, Rupak, and Savage]{Bedaque:2003wj}
Bedaque, P.F.; Rupak, G.; Savage, M.J.
\newblock {Goldstone bosons in the 3P(Z) superfluid phase of neutron matter and
  neutrino emission}.
\newblock {\em Phys. Rev. C} {\bf 2003}, {\em 68},~065802,
\newblock
  doi:{\changeurlcolor{black}\href{https://doi.org/10.1103/PhysRevC.68.065802}{\detokenize{10.1103/PhysRevC.68.065802}}}.

\bibitem[Tolos and Fabbietti(2020)]{Tolos:2020aln}
Tolos, L.; Fabbietti, L.
\newblock {Strangeness in Nuclei and Neutron Stars}.
\newblock {\em Prog. Part. Nucl. Phys.} {\bf 2020}, {\em 112},~103770,

\bibitem[Akmal \em{et~al.}(1998)Akmal, Pandharipande, and
  Ravenhall]{Akmal:1998cf}
Akmal, A.; Pandharipande, V.; Ravenhall, D.
\newblock {The Equation of state of nucleon matter and neutron star structure}.
\newblock {\em Phys. Rev. C} {\bf 1998}, {\em 58},~1804--1828,
\newblock
  doi:{\changeurlcolor{black}\href{https://doi.org/10.1103/PhysRevC.58.1804}{\detokenize{10.1103/PhysRevC.58.1804}}}.

\bibitem[Heiselberg and Hjorth-Jensen(2000)]{Heiselberg:1999mq}
Heiselberg, H.; Hjorth-Jensen, M.
\newblock {Phases of dense matter in neutron stars}.
\newblock {\em Phys. Rept.} {\bf 2000}, {\em 328},~237--327,
\newblock
  doi:{\changeurlcolor{black}\href{https://doi.org/10.1016/S0370-1573(99)00110-6}{\detokenize{10.1016/S0370-1573(99)00110-6}}}.

\bibitem[Tolos \em{et~al.}(2016)Tolos, Manuel, Sarkar, and
  Tarrus]{Tolos:2014wla}
Tolos, L.; Manuel, C.; Sarkar, S.; Tarrus, J.
\newblock {Transport coefficients in superfluid neutron stars}.
\newblock {\em AIP Conf. Proc.} {\bf 2016}, {\em 1701},~080001,
\newblock
  doi:{\changeurlcolor{black}\href{https://doi.org/10.1063/1.4938690}{\detokenize{10.1063/1.4938690}}}.

\bibitem[Andersson \em{et~al.}(2005)Andersson, Comer, and
  Glampedakis]{Andersson:2004aa}
Andersson, N.; Comer, G.; Glampedakis, K.
\newblock {How viscous is a superfluid neutron star core?}
\newblock {\em Nucl. Phys. A} {\bf 2005}, {\em 763},~212--229,
\newblock
  doi:{\changeurlcolor{black}\href{https://doi.org/10.1016/j.nuclphysa.2005.08.012}{\detokenize{10.1016/j.nuclphysa.2005.08.012}}}.

\bibitem[Manuel \em{et~al.}(2005)Manuel, Dobado, and
  Llanes-Estrada]{Manuel:2004iv}
Manuel, C.; Dobado, A.; Llanes-Estrada, F.J.
\newblock {Shear viscosity in a CFL quark star}.
\newblock {\em J. High Energy Phys.} {\bf 2005}, {\em 9},~76,
\newblock
  doi:{\changeurlcolor{black}\href{https://doi.org/10.1088/1126-6708/2005/09/076}{\detokenize{10.1088/1126-6708/2005/09/076}}}.

\bibitem[Alford \em{et~al.}(2010)Alford, Braby, and Mahmoodifar]{Alford:2009jm}
Alford, M.G.; Braby, M.; Mahmoodifar, S.
\newblock {Shear viscosity due to kaon condensation in color-flavor locked
  quark matter}.
\newblock {\em Phys. Rev. C} {\bf 2010}, {\em 81},~025202,
\newblock
  doi:{\changeurlcolor{black}\href{https://doi.org/10.1103/PhysRevC.81.025202}{\detokenize{10.1103/PhysRevC.81.025202}}}.

\bibitem[Rupak and Sch\"afer(2007)]{Rupak:2007vp}
Rupak, G.; Sch\"afer, T.
\newblock {Shear viscosity of a superfluid Fermi gas in the unitarity limit}.
\newblock {\em Phys. Rev. A} {\bf 2007}, {\em 76},~053607,
\newblock
  doi:{\changeurlcolor{black}\href{https://doi.org/10.1103/PhysRevA.76.053607}{\detokenize{10.1103/PhysRevA.76.053607}}}.

\bibitem[Mannarelli \em{et~al.}(2013)Mannarelli, Manuel, and
  Tolos]{Mannarelli:2012eg}
Mannarelli, M.; Manuel, C.; Tolos, L.
\newblock {Phonon contribution to the shear viscosity of a superfluid Fermi gas
  in the unitarity limit}.
\newblock {\em Ann. Phys.} {\bf 2013}, {\em 336},~12--35,
\newblock
  doi:{\changeurlcolor{black}\href{https://doi.org/10.1016/j.aop.2013.05.015}{\detokenize{10.1016/j.aop.2013.05.015}}}.

\bibitem[Bildsten and Ushomirsky(2000)]{Bildsten:1999zn}
Bildsten, L.; Ushomirsky, G.
\newblock {Viscous boundary layer damping of R modes in neutron stars}.
\newblock {\em Astrophys. J. Lett.} {\bf 2000}, {\em 529},~L33--L36,
\newblock
  doi:{\changeurlcolor{black}\href{https://doi.org/10.1086/312454}{\detokenize{10.1086/312454}}}.

\bibitem[Glampedakis and Andersson(2006)]{Glampedakis:2006mn}
Glampedakis, K.; Andersson, N.
\newblock {Ekman layer damping of r-modes revisited}.
\newblock {\em Mon. Not. R. Astron. Soc.} {\bf 2006}, {\em 371},~1311--1321,
\newblock
  doi:{\changeurlcolor{black}\href{https://doi.org/10.1111/j.1365-2966.2006.10749.x}{\detokenize{10.1111/j.1365-2966.2006.10749.x}}}.

\bibitem[Esel'Son \em{et~al.}(1980)Esel'Son, Nosovitskaya, Pogorelov, and
  Sobolev]{Eselson}
Esel'Son, B.N.; Nosovitskaya, O.S.; Pogorelov, L.A.; Sobolev, V.I.
\newblock {Characteristics of the viscosity of liquid helium below 1 K}.
\newblock {\em {Sov. J. Exp. Theor. Phys. Lett.}} {\bf 1980}, {\em 31},~34--37. 

\bibitem[Niemetz and Schoepe(2004)]{Niemetz2004}
Niemetz, M.; Schoepe, W.
\newblock {Stability of Laminar and Turbulent Flow of Superfluid 4 He at mK
  Temperatures Around an Oscillating Microsphere}.
\newblock {\em J. Low Temp. Phys.} {\bf 2004}, {\em
  135},~447--469,
\newblock
  doi:{\changeurlcolor{black}\href{https://doi.org/10.1023/B:JOLT.0000029507.98543.1d}{\detokenize{10.1023/B:JOLT.0000029507.98543.1d}}}.

\bibitem[Zadorozhko \em{et~al.}(2009)Zadorozhko, Rudavskii, Chagovets, Sheshin,
  and Kitsenko]{Zadorozhko}
Zadorozhko, A.A.; Rudavskii, E.Y.; Chagovets, V.K.; Sheshin, G.A.; Kitsenko,
  Y.A.
\newblock Viscosity and relaxation processes in the phonon-roton system of He
  II.
\newblock {\em Low Temp. Phys.} {\bf 2009}, {\em 35},~100--104,
\newblock
  doi:{\changeurlcolor{black}\href{https://doi.org/10.1063/1.3075937}{\detokenize{10.1063/1.3075937}}}.

\bibitem[Escobedo \em{et~al.}(2009)Escobedo, Mannarelli, and
  Manuel]{Escobedo:2009bh}
Escobedo, M.A.; Mannarelli, M.; Manuel, C.
\newblock {Bulk viscosities for cold Fermi superfluids close to the unitary
  limit}.
\newblock {\em Phys. Rev. A} {\bf 2009}, {\em 79},~063623,
\newblock
  doi:{\changeurlcolor{black}\href{https://doi.org/10.1103/PhysRevA.79.063623}{\detokenize{10.1103/PhysRevA.79.063623}}}.

\bibitem[Bierkandt and Manuel(2011)]{Bierkandt:2011zp}
Bierkandt, R.; Manuel, C.
\newblock {Bulk viscosity coefficients due to phonons and kaons in superfluid
  color-flavor locked quark matter}.
\newblock {\em Phys. Rev. D} {\bf 2011}, {\em 84},~023004,
\newblock
  doi:{\changeurlcolor{black}\href{https://doi.org/10.1103/PhysRevD.84.023004}{\detokenize{10.1103/PhysRevD.84.023004}}}.

\bibitem[Haensel \em{et~al.}(2000)Haensel, Levenfish, and
  Yakovlev]{Haensel:2000vz}
Haensel, P.; Levenfish, K.; Yakovlev, D.
\newblock {Bulk viscosity in superfluid neutron star cores. I. direct urca
  processes in npe mu matter}.
\newblock {\em Astron. Astrophys.} {\bf 2000}, {\em 357},~1157--1169.

\bibitem[Haensel \em{et~al.}(2001)Haensel, Levenfish, and
  Yakovlev]{Haensel:2001mw}
Haensel, P.; Levenfish, K.; Yakovlev, D.
\newblock {Bulk viscosity in superfluid neutron star cores. 2. Modified Urca
  processes in npe mu matter}.
\newblock {\em Astron. Astrophys.} {\bf 2001}, {\em 327},~130--137,
\newblock
  doi:{\changeurlcolor{black}\href{https://doi.org/10.1051/0004-6361:20010383}{\detokenize{10.1051/0004-6361:20010383}}}.

\bibitem[Andersson and Kokkotas(2001)]{Andersson:2000mf}
Andersson, N.; Kokkotas, K.D.
\newblock {The R mode instability in rotating neutron stars}.
\newblock {\em Int. J. Mod. Phys. D} {\bf 2001}, {\em 10},~381--442,
  doi:{\changeurlcolor{black}\href{https://doi.org/10.1142/S0218271801001062}{\detokenize{10.1142/S0218271801001062}}}.

\bibitem[Lindblom(2001)]{Lindblom:2000jw}
Lindblom, L.
\newblock {Neutron star pulsations and instabilities}.
\newblock {\em ICTP Lect. Notes Ser.} {\bf 2001}, {\em 3},~257--276.

\bibitem[Lindblom \em{et~al.}(1998)Lindblom, Owen, and
  Morsink]{Lindblom:1998wf}
Lindblom, L.; Owen, B.J.; Morsink, S.M.
\newblock {Gravitational radiation instability in hot young neutron stars}.
\newblock {\em Phys. Rev. Lett.} {\bf 1998}, {\em 80},~4843--4846,
\newblock
  doi:{\changeurlcolor{black}\href{https://doi.org/10.1103/PhysRevLett.80.4843}{\detokenize{10.1103/PhysRevLett.80.4843}}}.

\bibitem[Andersson \em{et~al.}(1999)Andersson, Kokkotas, and
  Schutz]{Andersson:1998ze}
Andersson, N.; Kokkotas, K.D.; Schutz, B.F.
\newblock {Gravitational radiation limit on the spin of young neutron stars}.
\newblock {\em Astrophys. J.} {\bf 1999}, {\em 510},~846,
\newblock
  doi:{\changeurlcolor{black}\href{https://doi.org/10.1086/306625}{\detokenize{10.1086/306625}}}.

\bibitem[Shternin and Yakovlev(2008)]{Shternin:2008es}
Shternin, P.; Yakovlev, D.
\newblock {Shear viscosity in neutron star cores}.
\newblock {\em Phys. Rev. D} {\bf 2008}, {\em 78},~063006,
\newblock
  doi:{\changeurlcolor{black}\href{https://doi.org/10.1103/PhysRevD.78.063006}{\detokenize{10.1103/PhysRevD.78.063006}}}.

\bibitem[Benhar and Carbone(2009)]{Benhar:2009nr}
Benhar, O.; Carbone, A.
\newblock {Shear viscosity of beta-stable nuclear matter}. \emph{arXiv} {\bf 2009},  arXiv:0912.0129.

\bibitem[Zhang \em{et~al.}(2010)Zhang, Lombardo, and Zuo]{Zhang:2010jf}
Zhang, H.F.; Lombardo, U.; Zuo, W.
\newblock {Transport parameters in neutron stars from in-medium NN cross
  sections}.
\newblock {\em Phys. Rev. C} {\bf 2010}, {\em 82},~015805,
\newblock
  doi:{\changeurlcolor{black}\href{https://doi.org/10.1103/PhysRevC.82.015805}{\detokenize{10.1103/PhysRevC.82.015805}}}.

\bibitem[Braby \em{et~al.}(2010)Braby, Chao, and Sch\"afer]{Braby:2009dw}
Braby, M.; Chao, J.; Sch\"afer, T.
\newblock {Thermal conductivity of color-flavor locked quark matter}.
\newblock {\em Phys. Rev. C} {\bf 2010}, {\em 81},~045205,
\newblock
  doi:{\changeurlcolor{black}\href{https://doi.org/10.1103/PhysRevC.81.045205}{\detokenize{10.1103/PhysRevC.81.045205}}}.

\bibitem[Shternin and Yakovlev(2007)]{Shternin:2007ee}
Shternin, P.; Yakovlev, D.
\newblock {Electron-muon heat conduction in neutron star cores via the exchange
  of transverse plasmons}.
\newblock {\em Phys. Rev. D} {\bf 2007}, {\em 75},~103004,
\newblock
  doi:{\changeurlcolor{black}\href{https://doi.org/10.1103/PhysRevD.75.103004}{\detokenize{10.1103/PhysRevD.75.103004}}}.

\bibitem[Bedaque and Reddy(2014)]{Bedaque:2013fja}
Bedaque, P.F.; Reddy, S.
\newblock \textls[-15]{{Goldstone modes in the neutron star core}. {\em Phys. Lett. B} {\bf 2014}, {\em 735},~340--343,  doi:{\changeurlcolor{black}\href{https://doi.org/10.1016/j.physletb.2014.06.033}{\detokenize{10.1016/j.physletb.2014.06.033}}}.}

\bibitem[Flowers \em{et~al.}(1976)Flowers, Ruderman, and
  Sutherland]{Flowers:1976ux}
Flowers, E.; Ruderman, M.; Sutherland, P.
\newblock {Neutrino pair emission from finite-temperature neutron superfluid
  and the cooling of young neutron stars}.
\newblock {\em Astrophys. J.} {\bf 1976}, {\em 205},~541,
\newblock
  doi:{\changeurlcolor{black}\href{https://doi.org/10.1086/154308}{\detokenize{10.1086/154308}}}.

\bibitem[Voskresensky and Senatorov(1987)]{Voskresensky:1987hm}
Voskresensky, D.; Senatorov, A.
\newblock {Description of Nuclear Interaction in Keldysh's Diagram Technique
  and Neutrino Luminosity of Neutron Stars.}
\newblock {\em Sov. J. Nucl. Phys.} {\bf 1987}, {\em 45},~411. (In Russian)

\bibitem[Yakovlev \em{et~al.}(1999)Yakovlev, Kaminker, and
  Levenfish]{Yakovlev:1998wr}
Yakovlev, D.; Kaminker, A.; Levenfish, K.
\newblock {Neutrino emission due to Cooper pairing of nucleons in cooling
  neutron stars}.
\newblock {\em Astron. Astrophys.} {\bf 1999}, {\em 343},~650.

\bibitem[Bedaque and Sen(2014)]{Bedaque:2013rya}
Bedaque, P.; Sen, S.
\newblock {Neutrino emissivity from Goldstone boson decay in magnetized neutron
  matter}.
\newblock {\em Phys. Rev. C} {\bf 2014}, {\em 89},~035808,
\newblock
  doi:{\changeurlcolor{black}\href{https://doi.org/10.1103/PhysRevC.89.035808}{\detokenize{10.1103/PhysRevC.89.035808}}}.

\bibitem[Kolomeitsev and Voskresensky(2008)]{Kolomeitsev:2008mc}
Kolomeitsev, E.E.; Voskresensky, D.N.
\newblock {Neutrino emission due to Cooper-pair recombination in neutron stars
  revisited}.
\newblock {\em Phys. Rev. C} {\bf 2008}, {\em 77},~065808,
\newblock
  doi:{\changeurlcolor{black}\href{https://doi.org/10.1103/PhysRevC.77.065808}{\detokenize{10.1103/PhysRevC.77.065808}}}.

\bibitem[Jaikumar \em{et~al.}(2002)Jaikumar, Prakash, and
  Sch\"afer]{Jaikumar:2002vg}
Jaikumar, P.; Prakash, M.; Sch\"afer, T.
\newblock {Neutrino emission from Goldstone modes in dense quark matter}.
\newblock {\em Phys. Rev. D} {\bf 2002}, {\em 66},~063003,
\newblock
  doi:{\changeurlcolor{black}\href{https://doi.org/10.1103/PhysRevD.66.063003}{\detokenize{10.1103/PhysRevD.66.063003}}}.

\bibitem[Nicolis and Penco(2018)]{Nicolis:2017eqo}
Nicolis, A.; Penco, R.
\newblock {Mutual Interactions of Phonons, Rotons, and Gravity}.
\newblock {\em Phys. Rev. B} {\bf 2018}, {\em 97},~134516.

\bibitem[Manuel and Llanes-Estrada(2007)]{Manuel:2007pz}
Manuel, C.; Llanes-Estrada, F.J.
\newblock {Bulk viscosity in a cold CFL superfluid}.
\newblock {\em J. Cosmol. Astropart. Phys.} {\bf 2007}, {\em 8},~1,
\newblock
  doi:{\changeurlcolor{black}\href{https://doi.org/10.1088/1475-7516/2007/08/001}{\detokenize{10.1088/1475-7516/2007/08/001}}}.

\bibitem[Mannarelli and Manuel(2008)]{Mannarelli:2008jq}
Mannarelli, M.; Manuel, C.
\newblock {Transport theory for cold relativistic superfluids from an analogue
  model of gravity}.
\newblock {\em Phys. Rev. D} {\bf 2008}, {\em 77},~103014,
\newblock
  doi:{\changeurlcolor{black}\href{https://doi.org/10.1103/PhysRevD.77.103014}{\detokenize{10.1103/PhysRevD.77.103014}}}.

\bibitem[Mannarelli \em{et~al.}(2020)Mannarelli, Grasso, Trabucco, and
  Chiofalo]{Mannarelli:2020ebs}
Mannarelli, M.; Grasso, D.; Trabucco, S.; Chiofalo, M.L.
\newblock {Hawking temperature and phonon emission in acoustic holes}. \emph{arXiv} {\bf 2020},  arXiv:2011.00019.

\bibitem[Volovik(2001)]{Volovik:2000ua}
Volovik, G.
\newblock {Superfluid analogies of cosmological phenomena}.
\newblock {\em Phys. Rept.} {\bf 2001}, {\em 351},~195--348,
  doi:{\changeurlcolor{black}\href{https://doi.org/10.1016/S0370-1573(00)00139-3}{\detokenize{10.1016/S0370-1573(00)00139-3}}}.

\bibitem[Esposito \em{et~al.}(2019)Esposito, Krichevsky, and
  Nicolis]{Esposito:2018sdc}
Esposito, A.; Krichevsky, R.; Nicolis, A.
\newblock {Gravitational Mass Carried by Sound Waves}.
\newblock {\em Phys. Rev. Lett.} {\bf 2019}, {\em 122},~084501,
\newblock
  doi:{\changeurlcolor{black}\href{https://doi.org/10.1103/PhysRevLett.122.084501}{\detokenize{10.1103/PhysRevLett.122.084501}}}.

\end{thebibliography}
\end{document}